\documentclass[%
 aip,
 jmp,%
 amsmath,amssymb,
preprint,%
]{revtex4-1}

\usepackage{graphicx}
\usepackage{dcolumn}
\usepackage{bm}
\usepackage{epstopdf}
\usepackage{color}
\usepackage{ulem}

\newcommand*{\ddp}{{\frac{\partial}{\partial p}}}

\newcommand*{\jk}{{j_1,\cdots,j_K}}
\newcommand*{\jkp}{{j_1,\cdots,j_k+1,\cdots,j_K}}
\newcommand*{\jkm}{{j_1,\cdots,j_k-1,\cdots,j_K}}

\begin{document}

\title{Real-Time and Imaginary-Time Quantum Hierarchal Fokker-Planck Equations}

\author{Yoshitaka TANIMURA}
\email{tanimura@kuchem.kyoto-u.ac.jp.}
\affiliation{Department of Chemistry, Graduate School of Science, Kyoto University, Kyoto 606-8502, Japan}
\date{\today}

\begin{abstract}
We consider a quantum mechanical system represented in phase space (referred to hereafter as ''Wigner space''), coupled to a harmonic oscillator bath. We derive quantum hierarchal Fokker-Planck (QHFP) equations not only in real time, but also in imaginary time, which represents an inverse temperature. This is an extension of a previous work, in which we studied a spin-boson system, to a Brownian system. It is shown that the QHFP in real time obtained from a correlated thermal equilibrium state of the total system possess the same form as those obtained from a factorized initial state. A modified terminator for the hierarchal equations of motion is introduced to treat the non-Markovian case more efficiently. Using the imaginary-time QHFP, numerous thermodynamic quantities, including the free energy, entropy, internal energy, heat capacity, and susceptibility can be evaluated for any potential. These equations allow us to treat non-Markovian, non-perturbative system-bath interactions at finite temperature. Through numerical integration of the real-time QHFP for a harmonic system, we obtain the equilibrium distributions, the auto-correlation function, and the first- and second-order response functions.  These results are compared with analytically exact results for the same quantities. This provides a critical test of the formalism for a non-factorized thermal state, and elucidates the roles of fluctuation, dissipation, non-Markovian effects, and system-bath coherence. Employing numerical solutions of the imaginary-time QHFP, we demonstrate the capability of this method to obtain thermodynamic quantities for any potential surface. It is shown that both types of QHFP equations can produce numerical results of any desired accuracy.  The FORTRAN source codes that we developed, which allow for the treatment of Wigner space dynamics with any potential form, (TanimuranFP15 and ImTanimuranFP15) are provided as supplementary materials.
\end{abstract}

                         

\maketitle

\section{INTRODUCTION}
A Brownian oscillator (BO) model, which consists of a primary system coupled to a harmonic oscillator bath, is a versatile model that has been used to investigate fundamental problems in physics, chemistry and biology.\cite{Feynman63, CaldeiraPRL81,CLAnnlPhys1983,KuboToda85, GrabertZPhys84, GrabertPR88,Weiss08, TanimuraJPSJ06} The key feature of the Brownian model is that it describes irreversible dynamics through which the system evolves toward the thermal equilibrium state at finite temperature. This feature arises from interaction with the heat bath, which exhibits the canonical distribution at temperature $T$. To make the heat bath an unlimited heat source that possesses infinite heat capacity, the number of heat bath oscillators is effectively made infinitely large by replacing the spectral distribution of the system-oscillator coupling, $J(\omega)$, which was originally defined as the discretized distribution $J(\omega)=\sum c_j^2 \delta(\omega-\omega_j)$ (where $c_j$ is the coupling strength between the system and the $j$th bath oscillator with frequency $\omega_j$), with a continuous distribution, for example, $J(\omega)\propto \omega$.
Because the time-evolution of the total system is described by the Schr\"odinger equation, the total energy is conserved and the dynamics are reversible. In the reduced description of the system obtained by tracing over the bath degrees of freedom using such methods as the path integral method\cite{Feynman63} or the projection operator method,\cite{KuboToda85,Weiss08} however, the energy is no longer conserved, and its dynamics are irreversible, because the reduced system is merely a part of the total system. 
Heat bath effects arise in the reduced dynamics as fluctuation and dissipation in the reduced main system. These satisfy the classical or quantum version of the fluctuation-dissipation theorem. The reduced system evolves in an irreversible manner toward the thermal equilibrium state, in which the energy supplied by the fluctuations and the energy lost through dissipation are balanced, while the bath temperature does not change, because its heat capacity is infinite.

With the above described features, the Brownian model exhibits wide applicability, despite its simplicity. This is because the influence of the environment can in many cases be approximated by a Gaussian process, due to the cumulative effect of the large number of weak environmental interactions, in which case the ordinary central limit theorem is applicable,\cite{Kampen81} while the distribution function of the harmonic oscillator bath itself also exhibits a Gaussian distribution. By adjusting the form of the spectral distribution, the properties of the bath can be adjusted to represent a variety of environments consisting of, for example, solid state materials, solvates, and protein molecules. This model has been used to solve various problems of practical interest, in particular to investigate tunneling processes,\cite{CaldeiraPRL81,CLAnnlPhys1983,WaxmanLeggett1985} chemical reaction,\cite{Wolyness1981,Miller1989} non-adiabatic transition,\cite{Garg86, Sparpagilione1988} quantum device systems,\cite{MasonHess89}  ratchet rectification,\cite{Hanggi97,Hanggi09} to evaluate the efficiency of  SQUID rings,\cite{Chen1986, Wellstood2008} and to analyze the line shapes in laser spectra.\cite{Mukamel95,TaniIshi09}  

While the Brownian model itself is fairly simple, it is somewhat difficult to apply in the quantum mechanical case not only analytically but also numerically, due to the infinite number of bath degrees of freedom. Analytically exact solutions of Green's function for the BO Hamiltonian have been obtained only in the cases of a harmonic oscillator,\cite{GrabertZPhys84,GrabertPR88} a free particle,\cite{HakimPRA85} and a free rotator,\cite{SuzukiJPSJ01} using the path integral approach. Several approximate approaches have been developed to facilitate application of the Brownian model to more complicated systems. These approaches involve variational methods to study polarons\cite{FeynmanHellwarth1962,Hellwarth1962} and the optical response of an anharmonic oscillator\cite{SuzukiOPT1999} using a damped oscillator as a trial function, an instanton method for estimating the tunneling rate using  instantaneously jumping paths between tunneling wells,\cite{CaldeiraPRL81,CLAnnlPhys1983} a WKB method for evaluating the density matrix along a classical minimal action path,\cite{WaxmanLeggett1985, Wolyness1981} and diagrammatic expansion methods to study the anharmonicity of potentials and the nonlinearity of the system-bath coupling.\cite{OkumraJCP96, OkumraJCP97,OkumraJPRE97}
The analytical expressions obtained in these studies are helpful to gain insight into the role of dissipative environments in the dynamics of systems, but they do not allow us to study situations investigated in modern experiments that are usually described by complex potentials driven with time-dependent external forces. 

A great deal of effort has been dedicated to numerically calculating the time evolution of BO systems under external perturbations. Widely used approaches employ a reduced equation of motion that can be derived from the quantum Liouville equation with the full Hamiltonian by reducing the heat bath degrees of freedom. To obtain reduced equations of motion in a compact form, one usually employs the Markovian assumption, in which the correlation time is very short in comparison to the characteristic time of the system dynamics. In this case, the noise can be regarded as white. The quantum Langevin equation and the quantum Fokker-Planck equation have been derived with the projection operator method and the path integral method, for example.\cite {CaldeiraPhysica83,ChenWaxman1985,Coffey07, Coffey07PCCP,Jyoti2011, Jyoti2014} 

In the classical case, the Langevin equation\cite{Langevinbook} and the Fokker-Planck (or Kramers) equation\cite{Kramers,Risken-Book} have proved to be useful in the treatment of transport problems, and they have even been included in algorisms employed in molecular dynamics simulations. However, the applicability of the quantum forms of these equations is very limited, because they cannot be derived in a quantum mechanical framework without severe approximations and/or assumptions. 
For example, in the treatment of the quantum Langevin equation expressed in operator form, it is generally assumed that the antisymmetric correlation function of the noise is very short (the Markovian assumption) and positive. A similar Markovian assumption has been used in the treatment of the quantum Fokker-Planck equation. But in order for these assumptions to be valid, the heat bath must be at a sufficiently high temperature, in which case most of the important quantum dynamical effects play a minor role. 
This implys that the Markovian assumption is incompatible with obtaining a quantum mechanical description of dissipative dynamics at low temperature.\cite{HermanGrabert}

An Ohmic spectral distribution is generally assumed to realize Markovian noise. 
As we show in Appendix B, however, even if the dissipation process is Markovian, the fluctuation process may not be, because it must satisfy the fluctuation-dissipation theorem.\cite{TanimuraJPSJ06} For this reason, if we apply the equation of motion under Markovian assumption to low temperature systems, then the positivity of the probability distributions of the reduced system cannot be maintained.\cite{Seragent} This is a fundamental limitation, known as the ''positivity problem,'' which is particularly significant for the quantum master equation.\cite{Davies76,Gorini78,Spohn80,Dumcke79,Pechukas94,Romero04, Frigerio81, Frigerio85}
If the system is not time dependent and if the system Hamiltonian and the system-bath interaction commute, the time-convolutionless (TCL) master equation becomes exact.\cite{TCLshibata77,TCLshibata79,TCLshibata10} For the time-dependent case and/or non-commuting case, however, this master equation is valid only to second order with respect to the system-bath interaction, and the positivity condition is again broken. As a method to preserve positivity, the rotating wave approximation (RWA), which modifies the interaction between the system and the heat bath, has been applied in order to put the equation of motion in the Lindblad form. However, the RWA alters the thermal equilibrium state and the dynamics of the reduced system. These changes are particularly large in the case of a strong system-bath coupling and at low temperature. 
Moreover, in a typical quantum transport problem, the system is described by continuous energy states, and the energy levels of the heat bath and the system overlap. For this reason, the RWA cannot be used. Treatments of these kinds are therefore not sufficient to construct fully quantum mechanical descriptions of broad validity.

Path integral Monte Carlo simulations do not have any of the limitations of the approaches discussed above, but this approach is computationally intensive, because the number of paths to be evaluated grows rapidly with time, while sampling fails, due to the phase cancellation of wave functions.\cite {EggerMak94,Makri95,CaoVoth96} Much effort has been made to overcome these problems and extend the applicability of this method.\cite{Makri96,Thorwart00,Makri07,JadhaoMakri08, Makri2D2010,Dattani12B, Dattani13, Sim2001,Dattani12A,Makri2014} Because this approach can easily incorporate the semi-classical approximation for the bath, it may be advantageous in the study of polyatomic systems treated in multi-dimensional coordinates, but applications to this point incorporating full quantum mechanical dynamics have been limited to relatively small systems without time-dependent external force.

Wave function based methodologies for the full Hamiltonian have been developed
in order to avoid the reduced description of the system.
 The multi-configurational time-dependent Hartree (MCTDH) approach\cite{ML-MCTDH1,ML-MCTDH2,WangTHoss1, WangTHoss2,WangTHoss3,WangTHoss4,ML-MCTTDH3,ML-MCTTDHRev} employs time-dependent basis sets to represent the total wave function. Then, a variational principle is applied to derive the optimal equation of motion in order to reduce the bath degrees of freedom. 
This approach can be used to treat nonlinear system-bath coupling and anharmonic bath modes.\cite{WangTHoss2} However, the number of bath modes must be increased until convergence is reached. This implies that the study of long time behavior requires more basis sets, which makes the calculation more difficult. In the effective-mode approach, the heat bath degrees of freedom are mapped to a linearly coupled harmonic oscillator chain. Then, the dynamics of the system are described by the wave function of the system with a finite number of chained oscillators using a truncation scheme\cite{Burghardt1,Burghardt2,Burghardt3} or by utilizing the density matrix renormalization group method.\cite{AlexPlenio} Strictly speaking, the time evolution obtained with the wave function based approach describes time-reversible processes and thus, within this approach, there exists no thermal equilibrium state. However, in practice, this kind of approach has wider applicability than the reduced equation of motion. At this stage, the results obtained from these approaches have been limited to relatively simple systems. In particular, the inclusion of time dependent external forces is not as straightforward in these approaches as in the case of reduced equation of motion, because the energy of the total system changes due to the presence of an external force if the perturbation is strong, and hence the optimal basis set may also be changed.

The reduced hierarchal equations of motion (HEOM), which are derived by differentiating the reduced density matrix elements defined by path integral, are reduced equations of motion that can describe the dynamics of the system for non-perturbative and non-Markovian system-bath interactions with any desired accuracy under strong time-dependent perturbations at finite temperature.\cite{TanimuraJPSJ06} In this formalism, the effects of higher-order non-Markovian system-bath interactions are mapped into the hierarchal elements of the reduced density matrix. In their original formulation, these equations of motion were limited to the case in which the spectral distribution function takes the Drude form (i.e., the Ohmic form with a Lorentzian cut-off) and the bath temperature is high.\cite{Tanimura89A} However, with the inclusion of low temperature corrections terms, this temperature limitation has been eliminated.\cite{TanimuraPRA90,IshizakiJPSJ05,Xu05,Yan06} In addition, with the extension of the dimension of the hierarchy, in its present form, this approach is capable of treating a great variety of spectral distribution functions.\cite{TanimuraMukamelJPSJ94,TanakaJPSJ09,TanakaJCP10,TanimruaJCP12,Nori12,KramerFMO, KramerJPC2013,YanBO12,Shi14} This formalism is valuable because it can be used to treat not only strong system-bath coupling but also quantum coherence between the system and bath, which is essential to study a system subject to a time-dependent external force\cite{TanimuraJPSJ06} and nonlinear response functions.\cite{IshizakiCP08,Tanimura89B, Tanimura89C}
The system-bath coherence becomes particularly important if the bath interaction is regarded as non-Markovian, as found from femtosecond nonlinear optical measurements, which are carried out on time scales that are much shorter than the noise correlation time of environmental molecules.\cite{Mukamel95} 

For a Brownian system, the reduced hierarchal equations of motion are expressed in the Wigner space representation.\cite{TanimuraPRA91,TanimuraJCP92,SteffenTanimura00,TanimuraSteffen00, KatoTanimura02,KatoTanimura04, SakuraiJPSJ13, SakuraiNJP14, SakuraiJPC11, KatoJPCB13, TanimuraMaruyama97, MaruyamaTanimura98,
 YaoYaoJCP14,GrossmannJPC14}
In the Markovian limit, these equations of motion reduce to the Caldeira-Leggett quantum Fokker-Planck equation,\cite{CaldeiraPhysica83,ChenWaxman1985} and in the classical limit, they reduce to the classical Fokker-Planck (Kramers) equation.\cite{Kramers, Risken-Book} 

Recently, the author derived the HEOM not only in real time, but also in imaginary time, which represents an inverse temperature, starting from correlated initial conditions for a system described by discretized energy states.\cite{Tanimura2014} 
Reduction of these HEOM to a system represented in Wigner space is not straightforward, because they involve derivatives with respect to the position  and momentum that require a careful treatment with regards to the order of time slices in the path integral formalism. In this paper, we present the derivation of real- and imaginary-time HEOM in Wigner space and demonstrate the validity of these equations. 

The organization of the paper is as follows. In Sec. II, we present a model Hamiltonian and its influence functional with correlated initial conditions. In Sec. III, we derive the real-time quantum hierarchal Fokker-Planck (the real-time QHFP) equations using the influence functional given in Sec. II. In Sec. IV, we derive the imaginary-time quantum hierarchal Fokker-Planck (the imaginary-time QHFP) equations, which are convenient for evaluating thermodynamic quantities of the system. In Sec. V, the validity of our approach is demonstrated 
through numerical integration of the real- and imaginary-time QHFP equations for a harmonic system and comparing the calculated results with the exact results obtained from analytical calculations. Section VI is devoted to concluding remarks.

\section{REDUCED HIERACHAL EQUATIONS OF MOTION FROM CORRELATED INITIAL CONDITIONS}
We consider the situation in which the system interacts with a heat bath that gives rise to dissipation and fluctuation in the system. To illustrate this, let us consider a Brownian Hamiltonian expressed as \cite{ Feynman63, CaldeiraPRL81,CLAnnlPhys1983,KuboToda85, GrabertZPhys84, GrabertPR88,Weiss08, TanimuraJPSJ06}
\begin{eqnarray}
  \hat{H}_{tot} =   \hat{H}_A (\hat{p}, \hat{q})
   + \sum_{j} \left[ \frac{\hat{p}_{j}^{2}}{2m_{j}} + \frac{m_{j}\omega_{j}^2}{2} \left( \hat{x}_{j} - \frac{\alpha_{j} \hat{q}}{m_{j}\omega_{j}^2} \right)^2 \right],
 \label{CLHamiltonian}
\label{eq:BrownianH}
\end{eqnarray}
where
\begin{eqnarray}
 \hat{H}_A (\hat{p}, \hat{q})= \frac{\hat{p}^2}{2m} + U(\hat{q})
\label{eq:SystemH}
\end{eqnarray}
is the Hamiltonian for the system with mass $m$ and potential $U(\hat q)$ described by the momentum $\hat{p}$ and position $\hat{q}$. The bath degrees of freedom are treated as an ensemble of harmonic oscillators, and the momentum, position, mass, and frequency of the $j$th bath oscillator are given by $\hat{p}_{j}$, $\hat{x}_{j}$, $m_{j}$ and $\omega_{j}$, respectively. In the conventional Brownian model, the system-bath interaction is represented by a bilinear function of the system and bath coordinates as $H_I=-{\hat q}\sum_{j} \alpha_{j} \hat x_{j} $. Brownian models employing this bilinear interaction have been studied with various approaches.\cite{KuboToda85, GrabertZPhys84,GrabertPR88,Weiss08} In this paper also we restrict our investigation to this bilinear form to simplify the derivation, but we note that extension to the non-bilinear case is possible. \cite{TanimuraJPSJ06,SteffenTanimura00,TanimuraSteffen00,KatoTanimura02,KatoTanimura04, SakuraiJPC11} To maintain translational symmetry in the case of $U(\hat q)=0$, required to describe 
the motion of a free Brownian particle, we include the counter-term $\sum\nolimits_{j} {\alpha_{j}^2  {\hat q}^2/2m_{j} \omega _{j}^2 }$ in Eq.\eqref{eq:BrownianH}.

The heat bath we consider is characterized by the spectral distribution function defined by $
  J(\omega) \equiv \sum_{j }
({\hbar \alpha_{j}^2}/{2m_{j}\omega_{j}}) 
\delta(\omega-\omega_{j})$
and the inverse temperature, $\beta \equiv 1/k_{\mathrm{B}}T$, where $k_\mathrm{B}$ is the Boltzmann constant. 
The path integral used here to derive the HEOM is expressed in terms of an influence functional with correlated initial conditions. The influence functional that we employ, $ F_{CI}[t, \beta\hbar]$, is calculated by taking the trace over the heat bath degrees of freedom, starting from the thermal equilibrium state of the total Hamiltonian. The calculation of the influence functional for a heat bath consisting of harmonic oscillators is analogous to that of the generating functional for a Brownian oscillator system if we regard the system operator in the system-bath interaction $\hat q$ as an external force acting on the bath.\cite{Tanimura93,OkumuraPRE96,TanimuraOkumuraJCP97}
As shown in Appendix A, the reduced density matrix elements of the system with correlated initial conditions can be expressed as
\begin{align}
\rho(q,q';t) =&  \frac1{Z_{tot}}\int_{q_0=q(0)}^{q=q(t)} D[q(t)] \int_{q_0'=q' (0)}^{q'=q' (t)} D[q'(t)] \int_{q_0=\bar q(0)}^{q_0'=\bar q(\beta \hbar)} D[\bar q(\tau)]\;
\nonumber \\
&\times {\rm e}^{\frac{i}{\hbar}S_A[q,t]}
F_{CI}[q, q', \bar q;\; t, \beta\hbar] \bar \rho ^{eq} [\bar q ; \beta\hbar] 
{\rm e}^{-\frac{i}{\hbar}S_A[q',t]}, 
\label{rho_red}
\end{align}
where $S_A [q;\,t] $ is the action for the Hamiltonian of the system, Eq. \eqref{eq:SystemH}, given by
\begin{eqnarray}
S_A [q;\,t] =  \int_{0}^{t} d\tau \left\{ \frac{1}{2} m \dot q^2(\tau)   - U(q (\tau)) \right\},
\label{eq:actionfunc}
\end{eqnarray}
and $\bar \rho^{eq} [\bar q; \beta \hbar]$ is the initial thermal equilibrium state, with the heat bath defined by Eq.\eqref{eq:PCI}.

We assume that the spectral density $J(\omega)$ has an Ohmic form
with a Lorentzian cut-off and write\cite{TanimuraJPSJ06}
\begin{equation}
 J(\omega) = \frac{\hbar m\zeta}{\pi}\frac{\gamma^2\omega}{\gamma^2+\omega^2},
\label{JDrude}
\end{equation}
where the constant $\gamma$ represents the width of the spectral distribution of the collective bath modes
and is the reciprocal of the correlation time of the noise induced by the bath. 
The parameter $\zeta$ is the system-bath coupling strength, which represents the magnitude of damping. This spectral distribution approaches the Ohmic distribution, $J(\omega)\approx \hbar m \zeta \omega /\pi$, for large $\gamma$. In Appendix B, we present several profiles of fluctuation and dissipation terms for the Drude distribution to illustrate the origin of the positivity problem in the Markovian master and Redfield equations.

With $J(\omega)$ given by Eq. \eqref{JDrude}, the influence functional with correlated initial conditions is expressed as\cite{Tanimura2014}
\begin{eqnarray}
F_{CI}[q, q', \bar q;\; t, \beta\hbar] &&= {\rm e}^{-
  \int_{0}^t dt'' \operatorname{e} ^{ - \gamma t'' }   
\Phi(t'' ) \left\{ \int_{0}^{t''} dt'  
  \operatorname{e} ^{  \gamma  t' } \gamma \Theta_0 (t' ) 
 + G_0(0)
-\frac1{\hbar} \bar \Theta  (\beta \hbar)  \right\}} \nonumber \\
&&\times {\rm e}^{ -
\int_{0}^t dt''\sum\limits_{k= 1}^{K} \operatorname{e} ^{ - \nu_k t'' }  
\Phi(t'' )\left\{ \int_{0}^{t''} dt' 
  \operatorname{e} ^{  \nu_k  t' } \nu_k \Theta_k (t' ) 
-\frac1{\hbar} \bar \Psi_k   (\beta \hbar) \right\}
 -  \int_{0}^t dt'' \Xi(t'')},
\label{eq:influenceFcoh}
\end{eqnarray}
where, for the Matsubara frequency $\nu_k \equiv 2\pi k/\beta \hbar$, we have defined
\begin{align}
\Phi (t) \equiv \frac{i}{\hbar }\left[ q(t)-q'(t)\right],
\label{eq:Phi}
\end{align}
\begin{eqnarray}
\Theta _0 (t) \equiv \frac{ m \zeta }{2} \left\{ 
\left[ \dot q(t)+ \dot q'(t)\right]
- i\gamma \cot \left( {\frac{{\beta \hbar \gamma }}{2}} \right)
\left[ q(t)-q'(t)\right] \right\},
\label{eq:Theta0}
\end{eqnarray}
\begin{align}
G_0(0) \equiv \frac{{ m \zeta \gamma }}{2} \left[ q(0)+q'(0)\right],
\label{eq:G0}
\end{align}
\begin{eqnarray}
\bar \Theta (\beta \hbar) \equiv 
\frac{2m\zeta \gamma^2}{\beta}\int_{0}^{\beta\hbar} d \tau'  \bar q(\tau')
\left\{\frac1{2\gamma}+\sum_{k=1}^{\infty} 
{\frac{\left[\gamma \cos (\nu_k \tau' ) - i \nu_k \sin (\nu_k \tau' ) \right]
}{{\gamma^2  - \nu_k^2 }}}
\right\},
\label{eq:Thetabar}
\end{eqnarray}
and for $k \ge 1$,
\begin{eqnarray}
\Theta_k (t) \equiv -\frac{i}{\hbar} 
 \frac{2 m\zeta \gamma ^2 }{\beta}
\frac{1}{{ \nu_k^2-\gamma^2   }} \left[ q(t)-q'(t)\right],
\label{eq:Theta_k}
\end{eqnarray}
\begin{eqnarray}
\bar \Psi_k  (\beta \hbar) \equiv -
\frac{2 m\zeta \gamma ^2 }{\beta}\int_{0}^{\beta\hbar} d \tau'  \bar q(\tau')
\frac{ \nu_k \left[ \cos(\nu_k \tau' )- i \sin (\nu_k \tau' ) \right] }{{\gamma^2  - \nu_k^2 }},
\label{eq:Psi_kbar}
\end{eqnarray}
and
\begin{align}
\Xi'(t) 
=-\frac{{m \zeta }}{\beta }\left[\sum_{k=K+1}^{\infty} \frac{2\gamma^2}{\gamma^2-\nu_k^2} C_k \right] \left[ q(t)-q'(t)\right]^2,
\label{eq:Xi}
\end{align}
where $ C_k \equiv {\nu_k^2}/({\nu_k^2+\omega_c^2})$ is the correction factor that counteracts the overestimation of the contribution of higher-order Matsubara frequencies approximated by the delta function with cut-off number, $K$, introduced in Appendix B for the characteristic frequency of the system, $\omega_c$. This modification improves the convergence of hierarchies at lower temperature.
We now introduce the hierarchal elements that play an essential role in our formalism:
\begin{eqnarray}
 \rho_{j_1,\dots,j_K}^{(n)}(q,q';t) &&= \frac{1}{Z_{tot}}
\int_{q_0=q(0)}^{q=q(t)} D[q(t)] \int_{q_0'=q' (0)}^{q'=q' (t)} D[q'(t)] \int_{q_0=\bar q(0)}^{q_0'=\bar q(\beta \hbar)} D[\bar q(\tau)] 
 \nonumber \\  
 && \times 
{\rm e}^{\frac{i}{\hbar}S_A[q,t]} 
F_{j_1, \cdots,j_K}^{(n)}[ q, q', \bar q; t, \beta\hbar] \bar \rho^{eq} [\bar q; \beta\hbar]
{\rm e}^{-\frac{i}{\hbar}S_A[q',t]},
\label{eq:rhon_jk}
\end{eqnarray}
where
\begin{eqnarray}
F_{j_1, \cdots,j_K}^{(n)}[q, q', \bar q;\; t, \beta\hbar] &&= 
 \left\{ \operatorname{e} ^{ -\gamma t } \left[
\int_{0}^{t} dt' 
  \operatorname{e} ^{ \gamma t' }\gamma \Theta_0 (t' ) 
 +  G_0(0) -\frac1{\hbar} \bar \Theta (\beta \hbar) \right] \right\} ^{n} \nonumber \\
&& \times \prod_{k=1}^K 
\left\{ \operatorname{e} ^{ -\nu_k  t} \left[
\int_{0}^{t} dt' 
  \operatorname{e} ^{ \nu_k t'} \nu_k  \Theta_k (t' ) 
-\frac1{\hbar}  \bar \Psi_k  (\beta \hbar) \right] \right\} ^{j_k} 
\nonumber \\&&\times 
F_{CI}[q, q', \bar q;\; t, \beta\hbar],
\label{auxFV}
\end{eqnarray}
for nonnegative integers $n,j_1,\dots,j_K$. From the above definition, the first hierarchal element and the reduced density matrix given by Eq.\eqref{rho_red} are identical: $\rho(q,q';t) = \rho_{0,\dots,0}^{(0)}(q,q';t)$. As shown in Appendix C, we then have the following equations of motion: 
\begin{align}
\frac{{\partial \rho _{{j_1}, \ldots ,{j_K}}^{\left( n \right)}\left( {q,q';t} \right)}}{{\partial t}} &=  - \left[ {\frac{i}{\hbar }L\left( {q,q'} \right) + n\gamma  + \sum\limits_{k = 1}^K {{j_k}{\nu _k} + \Xi' (q,q')} } \right]\rho _{{j_1}, \ldots ,{j_K}}^{\left( n \right)}\left( {q,q';t} \right) \nonumber \\
 &- n \gamma {{\bar \Theta }_0}(q,q')\rho _{{j_1}, \ldots ,{j_K}}^{\left( {n - 1} \right)}\left( {q,q';t} \right) ) \nonumber \\
 &- \sum\limits_{k = 1}^K {{j_k \nu_k}{\Theta _k}(q,q')\rho _{{j_1}, \ldots ,{j_k} - 1, \ldots ,{j_K}}^{\left( n \right)}\left( {q,q';t} \right)} \nonumber \\
 &- \Phi (q,q')\left( {\rho _{{j_1}, \ldots ,{j_K}}^{\left( {n + 1} \right)}\left( {q,q';t} \right) + \sum\limits_{k = 1}^K {\rho _{{j_1}, \ldots ,{j_k} + 1, \ldots ,{j_K}}^{\left( n \right)}\left( {q,q';t} \right)} } \right),
\label{eq:HEOMq}
\end{align}
where 
\begin{align}
L\left( {q,q'} \right) =  - \frac{{{\hbar ^2}}}{{2m}}\frac{{{\partial ^2}}}{{\partial {q^2}}} + \frac{{{\hbar ^2}}}{{2m}}\frac{{{\partial ^2}}}{{\partial {{q'}^2}}} + U(q) - U(q'),
\end{align}
\begin{align}
{{\bar \Theta }_0}(q,q') =   \frac{{i\hbar \zeta }}{{2}}&\left[ {\left( {\frac{\partial }{{\partial q}} - \frac{\partial }{{\partial q'}}} \right)}  {\  + \frac{{m\gamma }}{\hbar }\cot \left( {\frac{{\beta \hbar \gamma }}{2}} \right)\left( {q - q'} \right)} \right],
\end{align}
and $\Phi (q,q')$, ${\Theta _k}(q,q')$, and $\Xi' (q,q')$ are defined by Eqs.\eqref{eq:Phi},\eqref{eq:Theta_k}, and \eqref{eq:Xi} by making the replacements $q(t)\rightarrow q$ and $q'(t) \rightarrow q'$. 
In the HEOM formalism, only the first element $\rho(q,q';t) = \rho_{0,\dots,0}^{(0)}(q,q';t)$ has a physical meaning and the other elements $\rho_{{j_1}, \ldots ,{j_K}}^{( n )}(q,q';t)$ are introduced in numerical calculations in order to treat the non-perturbative and non-Markovian system-bath interaction. We can evaluate $\rho_{0,\dots,0}^{(0)}(q,q';t)$ through numerical integration of the above equations. 

We next explain the truncation scheme that we use for the hierarchical equations, which is different from the scheme used in previous studies.\cite{SakuraiJPSJ13, SakuraiNJP14,SakuraiJPC11, KatoJPCB13}
First, we choose the number of Matsubara frequencies to be included in the HEOM, $K$, such that it satisfies $K \gg \omega_0 /\nu_1$. Then, we introduce the scaled integer $K_{\gamma}$ as $K_{\gamma}\equiv$ int$( K \nu_1/\gamma )$ for $\nu_1>\gamma$ and $K_{\gamma}\equiv K$ for $\nu_1\le \gamma$, which allows us to make calculations in the highly non-Markovian case more efficiently. The index for the hierarchy, denoted by $n$, for a given value of $\gamma$, then runs from 0 to $K_{\gamma}$. The total number of hierarchy members to be included in the calculations is then given by $N\equiv(K_{\gamma}+K+1)!/(K+1)!/K_{\gamma}!$. For the case $\sum\nolimits_{k = 1}^K {{j_k}} > K$, we truncate the hierarchal equations by replacing Eq.\eqref{eq:HEOMq} with 
\begin{equation}
 \frac{\partial}{\partial t}{\rho_\jk^{(n)}}(q,q';t) = -\left({\hat L} + \hat{\Xi'} \right) \rho_\jk^{(n)}(q,q';t).
\label{rho_term}
\end{equation}
In practice, we can simply set $\rho _{{j_1}, \ldots ,{j_K}}^{\left( n \right)}\left( {q,q';t} \right) = 0$ instead of employing the above equation, because $\rho _{{j_1}, \ldots ,{j_K}}^{\left( n \right)}\left( {q,q';t} \right)$ decays to zero as $t$ becomes large.\cite{IshizakiJPSJ05}
For the $ K_{\gamma}$ and $ K_{\gamma}+1$ members of the hierarchy, we have the following relation, valid to order $\delta t$:
\begin{align}
\rho_{0, \ldots ,0}^{\left( {K_{\gamma} + 1} \right)}\left( {q,q';t} \right) &\simeq {\gamma ^{ - 1}}\left\{ { - \gamma {{\bar \Theta }_0}(q,q')\rho _{0, \ldots ,0}^{\left( K_{\gamma} \right)}(q,q';t) } \right.\nonumber \\
&\quad \left. { - \frac{1}{{N + 1}}\Phi (q,q')\left[ {\rho _{0, \ldots ,0}^{\left( {K_{\gamma} + 2} \right)}(q,q';t) - \sum\limits_{k = 1}^K {\rho _{0, \ldots 010 \ldots }^{\left( {K_{\gamma} + 1} \right)}(q,q';t)} } \right]} \right\}\nonumber \\
& \simeq  - {{\bar \Theta }_0}(q,q')\rho _{0, \ldots ,0}^{\left( K_{\gamma} \right)}(q,q';t).
\end{align}
This asymptotic relation allows us to obtain the terminator for given $\gamma$ in the form 
\begin{align}
\frac{{\partial \rho _{0, \ldots ,0}^{\left( {K_{\gamma}} \right)}\left( {q,q';t} \right)}}{{\partial t}} &=  - \left[ {\frac{i}{\hbar }L\left( {q,q'} \right) + K_{\gamma} \gamma   -\Phi (q,q'){{\bar \Theta }_0}(q,q')+ \Xi' (q,q') } \right]\rho _{0, \ldots ,0}^{\left( K_{\gamma} \right)}\left( {q,q';t} \right) \nonumber \\
& - K_{\gamma}\gamma{{\bar \Theta }_0}(q,q')\rho _{0, \ldots ,0}^{\left( {K_{\gamma} - 1} \right)}\left( {q,q';t} \right) .
\label{eq:HEOMqterm}
\end{align}
This equation reduces to the quantum Fokker-Planck equation in the Markovian limit, i.e., the Ohmic distribution ($\gamma \rightarrow \infty$) with the high-temperature limit.\cite{ CaldeiraPhysica83,TanimuraPRA91}

While the terms $\bar \Theta$ and $\bar \Psi_k$ from the correlated initial state  do not appear in Eqs.\eqref{eq:HEOMq} and \eqref{eq:HEOMqterm}, they define the hierarchal elements for the correlated initial equilibrium state. \cite{Tanimura2014} To demonstrate this point, we consider the initial states of the density operators, obtained by setting $t=0$ in Eq.\eqref{eq:rhon_jk}: 
\begin{eqnarray}
\rho_{j_1,\dots,j_K}^{(n)} (q,q';0) =\sum_{m=0}^n  {n \choose m } \left( G_0(0)\right) ^{n-m} \bar \rho_{j_1,\dots,j_K}^{(m)} (q,q';0),
\label{eq:correlatedHEOMG}
\end{eqnarray}
where
\begin{eqnarray}
\bar \rho_{j_1,\dots,j_K}^{(m)}(q,q';0) =\frac{1}{Z_{A}} \int_{q_0=\bar q(0)}^{q_0'=\bar q(\beta \hbar)} D[\bar q(\tau)] \left(-\frac1{\hbar} \bar \Theta  (\beta \hbar) \right)^{m}   \prod_{k=1}^K 
\left(  -\frac1{\hbar}  \bar \Psi_k  (\beta \hbar) \right)^{j_k} 
\bar \rho [\bar q; \beta\hbar]
   \nonumber \\
\label{eq:correlatedHEOMeq}
\end{eqnarray}
are the equilibrium hierarchal elements.   
Here, $Z_A$, $Z_{tot}$, and $Z_B$ are the partition functions of the system, total system, and bath, respectively, related as ${Z_{A}}=Z_{tot}/Z_{B}$. We then have $\rho^{eq}[\bar q; \beta \hbar]= Z_{B} \bar \rho [\bar q; \beta \hbar ]$. It is important to note that the steady state of $\rho_{{j_1}, \ldots ,{j_K}}^{n} (t)$ for $n>0$ in Eq.\eqref{eq:HEOMq} is slightly shifted from the initial thermal equilibrium state as a result of the influence of the sum in Eq. \eqref{eq:correlatedHEOMG}. However, because $\rho _{{0}, \ldots ,{0}}^{(0)} (t)$ is not influenced by this effect, expectation values calculated using $\rho _{{0}, \ldots ,{0}}^{(0)} (t)$ does not change.

From the above definition, it is clear that the HEOM members at time $t=0$ represent a correlated initial state, while the zeroth member, $\rho _{{0}, \ldots ,{0}}^{(0)} (0)= \bar \rho [\bar q; \beta\hbar]$, involves the static correlations. In Fig. \ref{LFig}, the correlations responsible for the correlated initial state are represented by green arcs, and the static correlations are represented by red arcs. After the time evolution, the elements $\rho_{j_1,\dots,j_K}^{(n)} (q,q';t)$ describe the dynamical correlation, represented by the blue arcs and lines in Fig. \ref{LFig}.

\section{Real-Time quantum Hierarchal Fokker-Planck equations}
We now introduce the Wigner distribution function, 
which is the quantum analog of the classical distribution function 
in phase space.   For the density matrix element $\rho_{j_1,\dots,j_K}^{(n)} (q,\,q';t)$, 
this is defined as~\cite{Wigner32, KuboWig64, Wigner84, Frensley90}
\begin{eqnarray}
W_{j_1,\dots,j_K}^{(n)}(p,q;t)  \equiv  \frac{1}
{{2\pi \hbar }}\int_{ - \infty }^\infty  {dx} \mathop e\nolimits^{{\rm i}px/\hbar }  \rho_{j_1,\dots,j_K}^{(n)} \left( {q - \frac{x}
{2},\,q + \frac{x}
{2}};t \right).
\end{eqnarray}
The Wigner representation of the reduced density matrix defined in Eq.\eqref{rho_red}, $W(p,q;t)$, and the first member of the hierarchal elements are then identical: $W(p,q;t) = W_{0,\dots,0}^{(0)}(p,q;t)$.
 The Wigner distribution function is a real function, in contrast to the complex density matrix. In terms of the Wigner distribution, the quantum Liouvillian takes the form~\cite{Frensley90}
\begin{eqnarray}
 - \mathcal{\hat L}_{QM}W_{j_1,\dots,j_K}^{(n)} (p,\,q) \equiv  - \frac{p}{m}\frac{\partial }{{\partial q}}W_{j_1,\dots,j_K}^{(n)} (p,\,q) 
- \frac{1}{\hbar }\int_{ - \infty }^\infty {\frac{{dp'}}
{{2\pi \hbar }} U_W (p - p',\,q)} W_{j_1,\dots,j_K}^{(n)} (p',\,q), \nonumber \\
\label{eq:quantumLioiv}
\end{eqnarray}
where $U_W(p,\,q)$ is given by
\begin{eqnarray}
U_W (p,\,q) = 2 \int_{0}^\infty dx \sin 
\left( {\frac{px}{\hbar}} \right) 
\left\{ {U \left( {q + \frac{x}{2}} \right) - U\left(  {q - \frac{x}{2}} \right)} \right\}.
\label{eq:wigop}
\end{eqnarray}
The quantum Liouvillian can also be expressed as\cite{KuboWig64,Wigner84}
\begin{eqnarray}
 - \mathcal{\hat L}_{QM}W_{j_1,\dots,j_K}^{(n)} (p,\,q) = \left[ - \frac{p}{m}\frac{\partial }{{\partial q}} 
+ \frac{1}{i \hbar } \left\{ U \left(q -\frac{\hbar}{2i}\frac{\partial}{\partial p}\right) -  U \left(q +\frac{\hbar}{2i}\frac{\partial}{\partial p}\right) \right\} \right] W_{j_1,\dots,j_K}^{(n)} (p,\,q). \nonumber \\
\label{eq:WignerMoyal}
\end{eqnarray}
While the above expression is easier to integrate in the case that the potential is nearly harmonic, the expression in Eq.\eqref{eq:quantumLioiv} is numerically stable, and it can be applied with any form of potential, including an unbounded potential.

Using the Wigner distribution and quantum Liouvillian, the equations of motion appearing in Eq. \eqref{eq:HEOMq} can be expressed in the form of quantum hierarchal Fokker-Planck (QHFP) equations in real time as
\begin{align}
 \frac{\partial}{\partial t}{W}_\jk^{(n)}(p,q;t) &= -\left[ \mathcal{\hat L}_{QM}  + n\gamma + \sum_{k=1}^K j_k\nu_k + \hat{\Xi'} \right] W_\jk^{(n)}(p,q;t) \nonumber \\
	& + \hat{\Phi} \left[ W_\jk^{(n+1)}(p,q;t) + \sum_{k=1}^K W_\jkp^{(n)}(p,q;t) \right] \nonumber \\
	& + n\gamma {{\hat {\bar \Theta} }_0} W_\jk^{(n-1)}(p,q;t) 
\nonumber \\
	& + \sum_{k=1}^K j_k\nu_k \hat{\Theta}_k W_\jkm^{(n)}(p,q;t),
\label{heom_wig}
\end{align}
where
$\hat \Phi =  {\partial }/{{\partial p}}$,
\begin{align}
{\hat {\bar \Theta} }_0 &\equiv { \zeta }
\left[p + \frac{m\hbar \gamma}{2} \cot \left(
\frac{\beta \hbar \gamma}{2} \right) \frac{\partial}{\partial p} \right],
\end{align}
\begin{equation}
 \hat{\Theta}_k \equiv -\frac{2m\gamma^2\zeta}{\beta (\nu_k^2-\gamma^2)}\ddp,
\end{equation}
and
\begin{equation}
 \hat{\Xi'} \equiv -\frac{m\zeta}{\beta}\left[\sum_{k=K+1}^{\infty} \frac{2\gamma^2}{\gamma^2-\nu_k^2} C_k  \right]\frac{\partial^2 }{{\partial p^2}}.
\end{equation}
As in the case of the energy eigenstate representation,\cite{Tanimura2014} the above equations are identical to the equations derived from factorized initial conditions.\cite{SakuraiJPSJ13, SakuraiNJP14,SakuraiJPC11, KatoJPCB13} The above equations are then truncated by using the modified ''terminators'' expressed in the Wigner representation. As explained in Sec. II, the number of Matsubara frequencies to be included in the calculation, $K$, is chosen to satisfy $K \gg \omega_c /\nu_1$. The upper limit for the number of hierarchy members for given  $\gamma$ is then chosen to be $K_{\gamma}\equiv$ int$( K \nu_1/\gamma )$ for $\nu_1>\gamma$ and $K_{\gamma}\equiv K$ for $\nu_1\le \gamma$. 
Then, for the case $\sum\nolimits_{k = 1}^K {{j_k}} > K$, we truncate the hierarchal equations by replacing Eq. \eqref{heom_wig} with
\begin{equation}
 \frac{\partial}{\partial t}{W}_\jk^{(n)}(p,q;t) = -\left( \mathcal{\hat L}_{QM} + \hat{\Xi'} \right) W_\jk^{(n)}(p,q;t),
\label{wig_term}
\end{equation}
while, for the case $n=K_{\gamma}$ we employ
\begin{align}
 \frac{\partial}{\partial t}{W}_{0, \ldots ,0}^{(K_{\gamma})}(p,q;t) &= -\left[ \mathcal{\hat L}_{QM}  + K_{\gamma}\gamma - \hat \Phi {{\hat {\bar \Theta} }_0}+ \hat{\Xi'}\right] W_{0, \ldots ,0}^{(K_{\gamma})}(p,q;t)  - K_{\gamma} \gamma {{\hat {\bar \Theta} }_0} 
W_{0, \ldots ,0}^{(K_{\gamma}-1)}(p,q;t).
\label{heom_wig_term}
\end{align}
We can evaluate $W_{j_1,\dots,j_K}^{(n)}(p,q;t)$ through numerical integration of the above equations. While only the first element $W(p,q;t) \equiv W_{0,0,\cdots,0}^{(0)}(p,q;t)$ has a physical meaning and the other elements  $W_{j_1,\dots,j_K}^{(n)}(p,q;t)$  are initially introduced to avoid the explicit treatment of the inherent memory effects, it turns out, however, that these elements allow us to take into account the system-bath coherence,\cite{TanimuraJPSJ06} entanglement\cite{DijkstraPRL10, DijkstraJPSJ12, Nori12} and expectation values that include the bath operators as $\langle \hat H_I  \rangle \equiv -\langle \hat q \sum \alpha_j {\hat x}_j \rangle$.\cite{Tanimura2014}
The HEOM consist of an infinite number of equations, but they can be evaluated with the desired accuracy by depicting the asymptotic behavior of the hierarchal elements for different $K$ and using this to determine whether or not there are sufficiently many members in the hierarchy. Essentially, the error introduced by the truncation to be negligibly small when $K$ is sufficiently large.

The correlated initial equilibrium state defined by Eq.\eqref{eq:correlatedHEOMG} is expressed in the Wigner representation accordingly.
The correlated initial equilibrium state can be set in the HEOM formalism by running the HEOM program until all of the hierarchy elements reach the steady state and then use these elements as the initial state,\cite{TanimuraJPSJ06} or by integrating the imaginary-time HEOM that we discuss in the next section.\cite{Tanimura2014} In practice, the former approach is simpler, because it requires the real-time HEOM only. This approach has been used to set the correlated initial conditions of the HEOM derived from factorized initial conditions that are identical to those used with the present HEOM.

The HEOM in Wigner space is ideal for studying quantum transport systems, because it allows the treatment of continuous systems, utilizing open boundary conditions and periodic boundary conditions.\cite{SakuraiJPSJ13, SakuraiNJP14}
In addition, the formalism can accommodate the inclusion of an arbitrary time-dependent external field. \cite{TanimuraJCP92,TanimuraMaruyama97, MaruyamaTanimura98,KatoJPCB13}

In the Markovian limit, $\gamma \rightarrow \infty$,
which is taken after the high temperature limit, yielding the condition  $\beta \hbar \gamma \ll 1$, we have the quantum Fokker-Planck equation\cite{CLAnnlPhys1983,WaxmanLeggett1985}
\begin{eqnarray}
\frac{\partial }{{\partial t}}W^{(0)} (p,\,q;\,t) =  - \mathcal{\hat L}_{QM} W^{(0)} (p,\,q;\,t) + \zeta \frac{\partial}{\partial p}\left( {p + \frac{m} {\beta }\frac{\partial}{{\partial p}}} \right)
 W^{(0)} (p,\,q;\,t),
\label{eq:GWLLFokkerPlanck}
\end{eqnarray}
which is identical to the quantum master equation without the RWA.\cite{TanimuraJPSJ06}
Because we assume that the relation $\beta \hbar \gamma \ll 1$ is maintained while taking the limit $\gamma \to \infty$,
this equation cannot be applied to low-temperature systems, in which quantum effects play a major role. As in the case of the master equation without the RWA, the positivity of the population distribution, $P(q) = \int dp W(p,q;t)$, cannot be maintained if we apply this equation in the low temperature case.

The classical HEOM can be derived by taking $\hbar \to 0$.\cite{TanimuraPRA91,TanimuraJCP92} 
The Wigner distribution function reduces to the classical one in this limit. 
The classical equation of motion is helpful,
because knowing the classical limit allows us to identify the purely quantum 
mechanical effects.\cite{TanimuraJCP92,SakuraiJPC11, KatoJPCB13} 

\section{Imaginary-Time quantum hierarchal Fokker-Planck equations}
The equilibrium reduced density matrix has been evaluated with several approaches.\cite{CaoPRB2012, CaoJCP2012}
By applying the methodology developed in Ref. \onlinecite{Tanimura2014}, we can derive the quantum hierarchal Fokker-Planck (QHFP) equations in imaginary time. This allows us to calculate the thermal equilibrium distribution $W^{eq}(p,q)$  at inverse temperature $\beta\hbar$. Instead of the quantum Liouvillian, this equation involves the left-sided operators $H_A(\hat p, \hat q)$ and $\hat q$. While the Wigner transformations of these operators become complex operators,
the Wigner distribution $W^{eq}(p,q)$ is a real function. In order to make  the numerical calculations easier to carry out, we rewrite $ \hat A \hat \rho$ as $(\hat A \hat \rho + \hat \rho \hat A)/2$ to perform the Wigner transformation, where $\hat A$ is an arbitrary operator. Other than this, the derivation of the imaginary-time QHFP is parallel to that of the imaginary-time HEOM in the energy eigenstate representation.\cite{Tanimura2014}
By introducing the Winger distribution for imaginary time, ${\bar W}_{k^1,\dots,k^{m}}^{\,[m:l]}(p,q; \tau)$, which is defined by the Wigner transformation of the density operator that in path integral form is given by
\begin{eqnarray}
\bar \rho_{k^1,\dots,k^{m}}^{\,[m:l]} (q_0 ,q_0'; \tau) &&= \int_{\bar q (0)=q_0}^{\bar q (\tau)=q'_0} 
D[\bar q(\tau ) ]
  \prod_{g=1}^{m-l} \left( \int_0^{\tau}  d\tau_g  \cos(\nu_{k^g}\tau_g)   \bar q(\tau_g) \right) \nonumber \\
&&\times\prod_{g'=m-l+1}^{2m} \left( \int_0^{\tau}  d\tau_{g'}  \sin(\nu_{k^{g'}}\tau_{g'})  \bar q(\tau_{g'}) \right) \bar \rho[\bar q ,\bar q '; \tau ],
\label{eq:ImHEOMelm}
\end{eqnarray}
we obtain the imaginary-time QHFP equations  as
\begin{eqnarray}
\frac{\partial}{\partial \tau} 
{\bar W}_{k^1,\dots,k^{m}}^{\,[m:l]} (\tau)
&& = - \bar H_A {\bar W}_{k^1,\dots,k^{m}}^{\,[m:l]}  (\tau)  
 +  \frac{1}{\hbar}
\sum\limits_{k^{m+1} =0}^{K} \bar c_{k^{m+1}} \cos(\nu_{k^{m+1}} \tau)   
q {\bar W}_{k^1,\dots,k^{m+1}}^{\,[m+1:l]}  (\tau)
 \nonumber  \\
&&  +   \frac{1}{\hbar}
\sum\limits_{k^{m+1} =0}^{K} \bar c_{k^{m+1}} \sin(\nu_{k^{m+1}} \tau)   
q {\bar W}_{k^1,\dots,k^{m+1}}^{\,[m+1:l+1]}  (\tau) \nonumber  \\
&& 
+  \frac{1}{\hbar}\sum\limits_{h =1}^{m-l}\cos(\nu_{k^h} \tau)   q {\bar W}_{k^1,\dots,k^{h-1},k^{h+1},\dots,k^{m}}^{\,[m-1:l]}  (\tau)
 \nonumber  \\
&& +  \frac{1}{\hbar}\sum\limits_{h =m-l+1}^{m}\sin(\nu_{k^h} \tau)  q {\bar W}_{k^1,\dots,k^{h-1},k^{h+1},\dots,k^{m}}^{\,[m-1:l-1]}  (\tau),
\label{eq:ImHEOMWig}
\end{eqnarray}
where the factors $\bar c_k$ are expressed as
$\bar c_0 = {m \zeta \gamma}/{\beta}$ and $\bar c_k = {2 m \zeta \gamma^2}/{\beta}{(\gamma+\nu_k)}$,
for $k \ge 1$. We set ${\bar W}_{k^1,\dots,k^{m}}^{\,[m:l]} (\tau)=0$ for higher-order elements in hierarchy denoted by $m$ to truncate. The Euclidean Liouvillian is expressed as
\begin{eqnarray}
\bar H_A \bar W = \frac{1}{2m}\left( p^2 - \frac{\hbar^2}{4} \frac{\partial^2}{\partial^2 q}   \right) \bar W(p',\,q) + 
\frac{1}{\hbar}\int_{ - \infty }^\infty \frac{dp'}{2\pi \hbar } 
 \bar U' (p - p',\,q) \bar W(p',\,q), 
\label{eq:EucLiouvint}
\end{eqnarray}
with
\begin{eqnarray}
\bar U' (p,\,q) =  \int_{0}^\infty dx \sin 
\left( {\frac{px}{\hbar}} \right) 
\left\{ {U' \left( {q + \frac{x}{2}} \right) + U' \left(  {q - \frac{x}{2}} \right)} \right\},
\label{eq:imwigop}
\end{eqnarray}
for the potential, $U'(q)=U(q)+m\zeta \gamma q^2/2$, including  the counter-term.
This can also be expressed in differential form as
\begin{eqnarray}
\bar H_A = \frac{1}{2m}\left( p^2 - \frac{\hbar^2}{4} \frac{\partial^2}{\partial^2 q}   \right) + \frac{1}{2}\left[ U'\left( { q - \frac{\hbar }{{2\rm i}}\frac{\partial }{{\partial p}}} \right)
+ U'\left( { q + \frac{\hbar }{{2\rm i}}\frac{\partial }{{\partial p}}} \right)\right].
\label{eq:EucLiouv}
\end{eqnarray}
If the anharmonicity of the potential is small, the above expression is useful. 
The initial conditions $\rho^{[0:0]} (q, q)=1$ and  $\rho^{[0:0]} (q, q')=0$ for $q\ne q'$ are expressed as
$\bar W^{[0:0]} (p,q; 0) = 1/2\pi$. 
By integrating Eq. \eqref{eq:ImHEOMWig} from $\tau=0$ to $\tau=\beta\hbar$, we can evaluate the equilibrium distribution function $\bar W(p,q; \beta\hbar)$.

Once we obtain the equilibrium distribution, we can calculate the partition function employing the relation
\begin{eqnarray}
Z_{A}(\beta \hbar) = \int dp \int dq {\bar W}^{[0:0]}(p,q; \beta\hbar ).
\label{eq:ZHEOM}
\end{eqnarray}
This allows us to calculate the Helmholtz free energy, $F_A=-\ln(Z_A)/\beta$, the entropy, $S_A=k_B \beta^2 \partial F_A/ \partial \beta$, the internal energy, $U_A=-\partial \ln (Z_A)/ \partial \beta$, and the heat capacity, $C_A=- k_B \beta^2 \partial U_A/\partial \beta$ for any potential. If the system is subject to an external force $\Delta f(\hat p, \hat q)$, where $f(\hat p, \hat q)$ is any function of the momentum and position, $\hat p$ and $\hat q$, we can also calculate the susceptibility, $\chi_A=-(\partial F/\partial \Delta)$, from $Z_A$.

It should be noted
 that even if the potential is a function of time, we can calculate thermodynamic quantities as functions of time through $Z_A(\beta \hbar; t)$, assuming that the system reaches the thermal equilibrium state faster than the change of the potential. 

\section{Numerical Results}
In principle, the HEOM provide an asymptotic approach that allows us to calculate various physical quantities with any desired accuracy by adjusting the number of hierarchal elements. Here, we demonstrate the applicability and validity of the real-time and imaginary-time QHFP equations, by presenting the results obtained from numerical integrations of Eqs.\eqref{heom_wig}-\eqref{heom_wig_term} and Eqs. \eqref{eq:ImHEOMWig}-\eqref{eq:imwigop}.
For this purpose, we consider the harmonic potential
\begin{eqnarray}
 \hat{H}_A (\hat{p}, \hat{q})= \frac{\hat{p}^2}{2m} + \frac{1}{2}m\omega_0^2\hat{q}^2.
\label{eq:Harmonic}
\end{eqnarray}
From our numerical solutions of Eqs.\eqref{heom_wig}-\eqref{heom_wig_term}, we have computed the equilibrium distributions, the auto-correlation functions, the first- and second-order response functions and examined the roles of a non-factorized thermal state, and the roles of fluctuation, dissipation, and system-bath coherence. From those of Eqs. \eqref{eq:ImHEOMWig}-\eqref{eq:imwigop}, we have computed the equilibrium distributions and thermodynamic quantities. 
Below, we compare these results with the same quantities calculated from analytically exact expressions for the Brownian oscillator system\cite{GrabertZPhys84, GrabertPR88, Weiss08} and from the time-convolutionless (TCL) Redfield equation both with and without the rotating wave approximation (RWA) \cite{TCLshibata77,TCLshibata79,TCLshibata10} (see Appendix D) as critical non-perturbative and non-Markovian tests. Note that the TCL equation is exact if the system Hamiltonian is time independent and if the system Hamiltonian and the system-bath interaction commute. However, here we consider the non-commuting case.

Below we also present our results for calculations of thermodynamic quantities obtained from the imaginary-time QHFP and compared them with analytical results.

\begin{figure}
\begin{center}
\scalebox{0.5}{\includegraphics{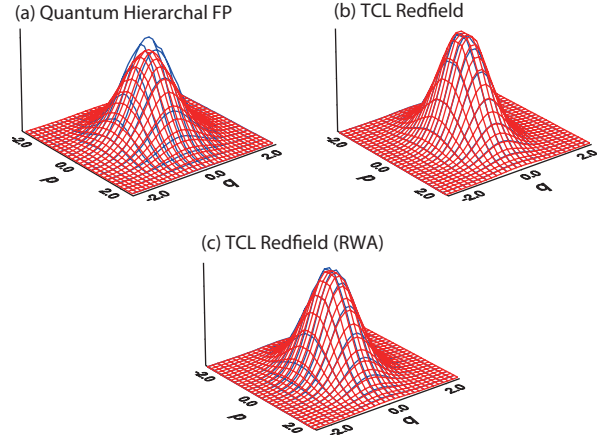}}
\end{center} 
\caption{\label{realWigner}(a) The initial conditions (blue curves) and steady state solutions (red curves) for the low temperature case $\beta\hbar=3.0$, calculated from (a) the real-time QHFP, (b) the TCL Redfield equation, and (c) the TCL Redfield equation with the RWA. The other parameter values are $\omega_0=1.0$, $\gamma=1.0$, and $\zeta=1.0$.
The factorized initial state given by $W_A^{eq}(p, q)$ with Eqs.\eqref{eq:QQHO} and \eqref{eq:PPHO} is set as the temporally initial state at time $t=0$. After integrating the real-time QHFP and the TCL Redfield equations for a sufficiently long time ($t=100$), the distribution reaches the steady state. In the real-time QHFP case, the obtained steady state is identical within numerical error to the thermal equilibrium state $W_{\rm BO}^{eq}(p, q)$ with $\langle q^2 \rangle$ and $\langle p^2 \rangle$ given by Eqs.\eqref{eq:QQBO} and \eqref{eq:PPBO}, while those from the TCL Redfield equations are similar to the original factorized initial state. This implies that the TCL Redfield equation cannot take into account the system-bath correlation properly.}
\end{figure}

\subsection{Steady state distribution: Static system-bath coherence and mixed state}
For a harmonic system, the equilibrium distribution in the Wigner representation is analytically expressed as
\begin{eqnarray}
W^{eq}(p, q) = \frac{1}{\bar N}\exp \left[-\frac{1}{2 \langle p^2 \rangle}p^2-\frac{1}{2\langle q^2 \rangle} q^2 \right],
\label{eq:WeqBO}
\end{eqnarray}
where $\bar N\equiv 2\pi\sqrt{\langle p^2 \rangle \langle q^2 \rangle}$ is the normalization factor and $\langle q^2 \rangle$ and $\langle p^2 \rangle$ are the mean squares of the position and momentum, respectively. 

The Wigner distribution for an isolated oscillator is written $W_A^{eq}(p, q)$. For the Hamiltonian Eq.\eqref{eq:Harmonic}, we have \cite{Wigner84}
\begin{eqnarray}
\langle q^2 \rangle_A = \frac{\hbar}{2m \omega_0 }\coth \left( \frac{\beta \hbar \omega_0}{2} \right)
\label{eq:QQHO}
\end{eqnarray}
and
\begin{eqnarray}
\langle p^2 \rangle_A = \frac{m \hbar \omega_0}{2}\coth \left( \frac{\beta \hbar \omega_0}{2} \right).
\label{eq:PPHO}
\end{eqnarray}
The Wigner distribution for a harmonic Brownian system is denoted by $W_{\rm BO}^{eq}(p, q)$. In this case, we have\cite{GrabertZPhys84, GrabertPR88, Weiss08}
\begin{eqnarray}
\langle q^2 \rangle_{\rm BO} = \frac{1}{m\beta} \sum_{k=-\infty}^{\infty} \frac{1}{\omega_0^2+\nu_k^2+ \left| \delta \Gamma^2 (\nu_k) \right| }
\label{eq:QQBO}
\end{eqnarray}
and 
\begin{eqnarray}
\langle p^2 \rangle_{\rm BO} =\frac{m}{\beta} \sum_{k=-\infty}^{\infty} \frac{\omega_0^2+\left|\delta \Gamma^2 (\nu_k)\right| }{\omega_0^2+\nu_k^2+\left|\delta \Gamma^2 (\nu_k)\right| },
\label{eq:PPBO}
\end{eqnarray}
with $\delta\Gamma^2(\omega)\equiv{\zeta \gamma^2 \omega}/({\gamma^2+\omega^2})$. In the Wigner representation, the thermal equilibrium state under the factorized assumption, $\exp(-\beta \hat H_A)\exp(-\beta \hat H_B )$, is denoted by $W_A^{eq}(p, q)$, while the true thermal equilibrium state of the reduced density operator, $tr_B\{\exp[-\beta (\hat H_{A}+\hat H_I+\hat H_B)]\}$, is denoted by $W_{\rm BO}^{eq}(p, q)$; the difference between the two distributions arises from the static system-bath coherence and represents the non-factorized effect of the thermal equilibrium state.
 
To obtain the thermal equilibrium state from the real-time QHFP, we integrated Eqs. \eqref{heom_wig}-\eqref{heom_wig_term} from a temporal initial state until all of the hierarchy elements reach the steady state. In principle, the initial state can have any form, but to elucidate the difference between the factorized (pure) equilibrium state and the true correlated (mixed) equilibrium state, we chose $W_{0,\dots,0}^{(0)} (p,q; 0)=W_A^{eq}(p, q) $ and $W_{j_1,\dots,j_K}^{(n)} (p,q;0) =0$ for other elements in the QHFP case.  For the TCL Redfield case, we chose $\rho_{jj} (0)=\exp(-\beta E_j' )/Z_A'$, where $E_j'$ is the $j$th eigenenergy of Eq.\eqref{eq:Harmonic} with the counter-term $\hat H_A' = (\hat H_A + m \zeta \gamma \hat q^2 /2)$, and $Z_A'=\sum_j \exp(-\beta E_j )$, as explained in Appendix D.

For all of our computations, we fixed the oscillator frequency as $\omega_0=1.0$. Then, we chose the coupling strength, inverse correlation time, and inverse temperature as $\zeta=1$, $\gamma=1$, and $\beta=3$. We thus consider the case of intermediate coupling strength and low temperature. For the QHFP, we set $K=7$, which leads to the depth in terms of $\gamma$ as $K_{\gamma} =2$ and the total number of hierarchy elements $N=4268$. The mesh size of the Winger function was optimized for the Liouvillian given in Eq. \eqref{eq:quantumLioiv},\cite{Frensley90} and we used $n_q =80$ and $n_p =30$ for the region $|q|<2.8$ . For the TCL Redfield equation, we employed six eigenstates. The calculated results and factorized initial state were translated into the Wigner distribution through Eqs. \eqref{eq:WigTCL} and \eqref{eq:WigTCLeq}, respectively.

In Fig.\ref{realWigner}(a), we display $W_{0,\dots,0}^{(0)} (p,q; t)$ for the factorized initial state at $t=0$  given by $W_A^{eq}(p, q)$ (blue curves) and the steady state distribution at $t=100$ (red curves) obtained from the real-time QHFP calculation. We found that even if we start from the factorized initial state, the steady state solution is the true thermal equilibrium state, denoted by $W_{\rm BO}^{eq}(p, q)$. This indicates that the real-time QHFP has the capability to produce the thermal equilibrium state with a static system-bath correlation through the fluctuation and dissipation terms. 
In the TCL Redfield equation cases, Figs. \ref{realWigner}(b) and (c), the calculated steady states (red curves) are similar to the factorized initial states (blue curves). The peak intensity of the TCL result in the case without the RWA is slightly higher than that in the case with the RWA, because the ground and first excited populations in the former case are $\rho_{00}(t)=1.083$ and $\rho_{11}(t)=-0.090$, due to the breakdown of the positivity condition, which is a physical requirement for the reduced equations of motion necessary for the population state of the density matrix to be positive. \cite{Davies76,Gorini78,Spohn80,Dumcke79,Pechukas94,Romero04, Frigerio81, Frigerio85} Other than this difference, the TCL distributions are similar to the factorized distribution. This indicates that the TCL Redfield equation cannot take into account the static system-bath coherence, because the TCL theory in the present case is valid only to second order in the system-bath coupling.

As explained in Appendix A, the effects of the system-bath coherence consist of the imaginary-time (static) part and complex-time (correlated) part represented by the red and green arcs in Fig. \ref{LFig}, respectively. From the equilibrium distribution, we can only observe the effects of the static part. To elucidate the correlated part, we need to calculate the nonlinear response function, as will be discussed in Sec. V-C.

\subsection{Two-body correlation functions: The roles of fluctuation, dissipation, and non-Markovian effects}

We next calculate the two-body correlation functions to investigate the roles of dissipation, fluctuation and non-Markovian dynamics.
The symmetric correlation and linear (first-order) response functions of the position are defined by $C(t)\equiv\langle \hat q(t) \hat q+ \hat q \hat q(t) \rangle/2$ and $R^{(1)}(t)\equiv\langle [\hat q(t), \hat q] \rangle/\hbar$, respectively. While the auto-correlation function of the position is given by $C(t)$, the observable of a linear measurement involving infrared and THz spectra, which are expressed in terms of a dipole proportional to $q$ corresponds to $R^{(1)}(t)$.
The Fourier transformation of these functions are denoted by $C[\omega]$ and $R^{(1)}[\omega]$. They are expressed as the real and imaginary parts of the normalized spectral distribution for the Brownian oscillator as\cite{, GrabertZPhys84,GrabertPR88}
\begin{eqnarray}
J' (\omega) = \frac{\hbar}{m}\frac{1}{(\omega_0^2-\omega^2)+i\omega I[i\omega] },
\end{eqnarray}
where $I[s]$ is the Laplace transformation of $B(t)$ defined by Eq. \eqref{eq:B_t} as
\begin{eqnarray}
I[s] = \int_{0}^{\infty} dt \frac{1}{m}B(t) \exp (-st).
\end{eqnarray}
For the Drude distribution, Eq.\eqref{JDrude}, we have 
$I[s]=\zeta \gamma/(s+\gamma)$ and
\begin{eqnarray}
C[\omega] = \frac{\hbar}{m}\frac{\delta\Gamma^2(\omega)\coth
\left( \frac{\beta\hbar\omega}{2}\right)}{\left(\omega^2-\omega_0^2-\delta\Omega^2(\omega)\right)^2+ \left(\delta\Gamma^2(\omega)\right)^2}
\label{eq:C_w}
\end{eqnarray}
and
\begin{eqnarray}
R^{(1)}[\omega] =\frac{\hbar}{m} \frac{\delta\Gamma^2(\omega)}{\left(\omega^2-\omega_0^2-\delta\Omega^2(\omega)\right)^2+ \left(\delta\Gamma^2(\omega)\right)^2},
\label{eq:R1_w}
\end{eqnarray}
where $\delta\Omega^2(\omega) \equiv {\zeta \gamma \omega^2}/({\gamma^2+\omega^2}) $.

In order to calculate the above functions using an equation of motion approach, we employ the following forms:\cite{TanimuraJPSJ06,TanimuraCP98}
\begin{eqnarray}
C(t)=\frac{1}{2}tr\left\{ \hat q \hat G(t) \hat q^{\circ} \hat \rho_{tot}^{eq}\right\}
\label{eq:C_L}
\end{eqnarray}
and
\begin{eqnarray}
R^{(1)}(t)=\frac{i}{\hbar}tr \left\{ \hat q \hat G(t) \hat q^{\times} \hat \rho_{tot}^{eq}\right\},
\label{eq:R1_L}
\end{eqnarray}
where  $\hat q^{\times} \hat A \equiv \hat q \hat A - \hat A \hat q$,$\hat q^{\circ} \hat A \equiv \hat q \hat A + \hat A \hat q$,
$\hat G(t) \hat A \equiv e^{-i \hat H_{tot} t/\hbar}\hat A e^{i \hat H_{tot} t/\hbar}$ for any operator $\hat A$, and $\hat \rho_{tot}^{eq}=e^{-\beta \hat H_{tot}}/Z_{tot}$ with $Z_{tot}=tr\{ \hat \rho_{tot}^{eq} \}$.

In the reduced equation of motion approach, the density matrix is replaced by a reduced one. In the QHFP case, $\hat \rho_{tot}^{eq}$ is replaced by the hierarchy member $W_{j_1,\dots,j_K}^{(n)} (p,q;t)$, whereas in the TCL Redfield case, it is replaced by $\hat \rho_{jk}(t)$. The Liouvillian in $\hat G(t)$ is replaced using Eqs. \eqref{heom_wig}-\eqref{heom_wig_term} and Eqs. \eqref{eq:RedfieldEq}-\eqref{eq:Redfieldel}, respectively.

We evaluate Eqs. \eqref{eq:R1_L} and \eqref{eq:C_L} in the following five steps.\cite{TanimuraJPSJ06,TanimuraCP98} (i) We first run the computational program to evaluate Eqs. \eqref{heom_wig}-\eqref{heom_wig_term} in the QHFP case and Eqs. \eqref{eq:RedfieldEq}-\eqref{eq:Redfieldel} in the TCL Redfield case for sufficiently long times from the temporal initial conditions to obtain a true thermal equilibrium state, as illustrated in Sec. V-A. In the QHFP case, the full hierarchy members $W_{j_1,\dots,j_K}^{(n)} (p,q; 0)$ are then used to set the correlated initial thermal equilibrium state. 
(ii) The system is excited by the first interaction $\hat q^{\times}$ or $\hat q^{\circ}$ at $t=0$. In the Wigner representation, they are expressed as $\partial/\partial p$ and $2q$, respectively. (iii) The evolution of the perturbed elements is then computed by running the program for the QHFP or TCL up to time $t$. (iv) Finally, the functions defined in Eqs.\eqref{eq:R1_L} and \eqref{eq:C_L} are calculated from the expectation value of $q$. By performing a fast Fourier transform, we obtain their spectra.

In computing the results reported below, we chose the number of Matsubara frequencies for the QHFP equation as $K = 5\text{--}8$, which leads to the depth in terms of $\gamma$ as $K_{\gamma} =3\text{--}6$ and the total number of hierarchy member $N=601\text{--}16093$. The mesh size of the Winger function was optimized for the Liouvillian given by Eq. \eqref{eq:quantumLioiv},\cite{Frensley90} and we used $n_q =80\text{--}120$ for the region $|q|<4\text{--}6$ and $n_p=30\text{--}120$ for the region $|p|<2.8\text{--}11.2$. In the TCL Redfield cases with and without the RWA, we varied the number of energy levels between 6 and 16 depending on the temperature.

\begin{figure}
\begin{center}
\scalebox{0.6}{\includegraphics{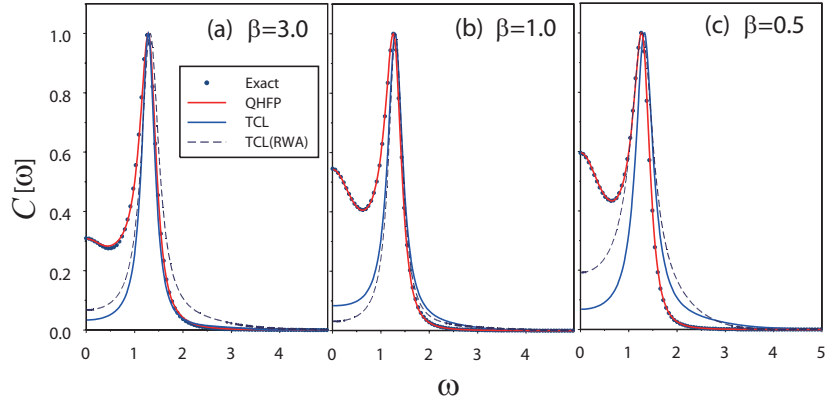}}
\end{center} 
\caption{\label{CSFig}The auto-correlation (symmetric correlation) function of the Brownian oscillator system for several inverse temperatures: (a) $\beta\hbar=3.0$, (b) $\beta\hbar=1.0$, (c) $\beta\hbar=0.5$. The dotted, red, blue, and blue-dash curves represent the results obtained from the analytic expression Eq. \eqref{eq:C_w}, the QHFP, the TCL-Redfield, and TCL-Redfield with the RWA, respectively. The intensity of each line is normalized with respect to its maximum peak intensity. The other parameters values are fixed as $\omega_0=1.0$, $\gamma=1.0$ and $\zeta=1$. } 
\end{figure}

\subsubsection{Auto-correlation function: Fluctuation and temperature effects}
First we study the temperature dependence of the auto-correlation function for the fixed coupling strength $\zeta=1$ and the inverse noise correlation time $\gamma=1$. 
In Fig. \ref{CSFig}, we compare the calculated real-time QHFP results obtained from Eqs.\eqref{heom_wig}\text{--}\eqref{heom_wig_term} with analytical results obtained  from Eq.\eqref{eq:C_w} and results obtained from the TCL Redfield equation, given in Eqs. \eqref{eq:RedfieldEq}-\eqref{eq:Redfieldel} without the RWA using Eq.\eqref{eq:NonRWA} and with the RWA using Eq. \eqref{eq:RWA}, for three values of the inverse temperature: (a) $\beta\hbar=3.0$, (b) $\beta\hbar=1.0$, (c) $\beta\hbar=0.5$.
At high temperature, in the QHFP case, the calculations are easier, because there are fewer Matsubara frequency terms, while the TCL Redfield calculations are more difficult, because more energy eigenstates are needed to account for the high energy excitations. Here we included up to 16 states in the TCL case.

While the QHFP results (red curves) coincide with the exact results (black dots), the TCL-Redfield results without the RWA (blue curves) and with the RWA (blue dashed curves) are close only near the maximum peak, regardless of temperature. The low-frequency parts of the spectra arise from the slow dynamics of the reduced system near the thermal equilibrium state, and the discrepancy between the TCL results and exact results arises from the equilibration process discussed in Sec. V-A. 

\begin{figure}
\begin{center}
\scalebox{0.6}{\includegraphics{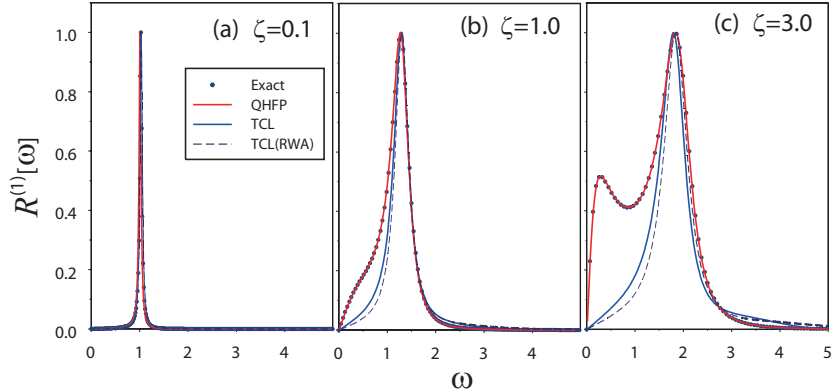}}
\end{center} 
\caption{\label{R1Fig}Linear response function for the Brownian oscillator system, $R^{(1)}[\omega]$, for three values of the system-bath coupling strengths: (a) $\zeta=0.1$, (b) $\zeta=1.0$, (c) $\zeta=3.0$. This function is temperature independent in the harmonic case, and we set the inverse temperature to $\beta\hbar =1$. The other parameter values are the same as in Fig. \ref{CSFig}. The dots represent the analytically calculated exact results obtained from Eq.\eqref{eq:R1_w}. The red, blue, and blue-dashed curves were calculated using the real-time QHFP equation, the TCL Redfield equation, and the TCL Redfield equation with the RWA, respectively. The intensity of each line is normalized with respect to its maximum peak strength.}
\end{figure}

\subsubsection{Linear response function: Dissipation and non-perturbative effects}
As can be seen from Eq.\eqref{eq:R1_w}, $R^{(1)}[\omega]$ is temperature independent. Therefore, this function is convenient to study the non-perturbative effects of the system-bath coupling, $\zeta$, and non-Markovian effects for slow modulation, controlled by the parameter $\gamma$, apart from the temperature effects. In Fig. \ref{R1Fig}, we compare the linear response functions for the coupling strengths (a) $\zeta=0.1$, (b) $\zeta=1.0$, and (c) $\zeta=3.0$ with fixed inverse temperature $\beta \hbar =1$ and $\gamma=1$.  

While the QHFP results (red curves) coincide with the exact results (black dots), the TCL-Redfield results without the RWA (blue curves) and with the RWA (the blue-dashed curves) are close only in the weak coupling case considered in Fig. \ref{R1Fig}(a). For the strong coupling case considered in Fig. \ref{R1Fig} (c), both the QHFP and analytical results exhibit a peak near $\omega_0=0.2$. This peak arises from the strong coupling between the harmonic mode and the low frequency bath mode characterized by $\gamma^2 \omega/(\gamma^2+\omega^2 )$ and only appears in the simultaneous non-Markovian ($\gamma \le \omega_0$) and non-perturbative ($\zeta \gg \omega_0$) case.\cite{SuzukiCP02} The existence of this peak, which we call a ``non-Markovian bosonic peak,`` is a good indication of the applicability of non-perturbative and non-Markovian theories.

Because the TCL Redfield theory is valid only to second order in the system-bath coupling, the TCL results cannot reproduce this peak. Moreover, the spectrum calculated from the TCL Redfield equation without the RWA in the strong coupling case, shown in Fig. \ref{R1Fig}(c), is not positive for $\omega \approx 5$, due to the breakdown of the positivity condition.
Despite this problem, however, the difference between the TCL results with and without the RWA is minor. This is because the spurious behavior caused by the positivity problem is suppressed in the non-Markovian treatment of the reduced dynamics, as explained in Appendix B.

\begin{figure}
\begin{center}
\scalebox{0.6}{\includegraphics{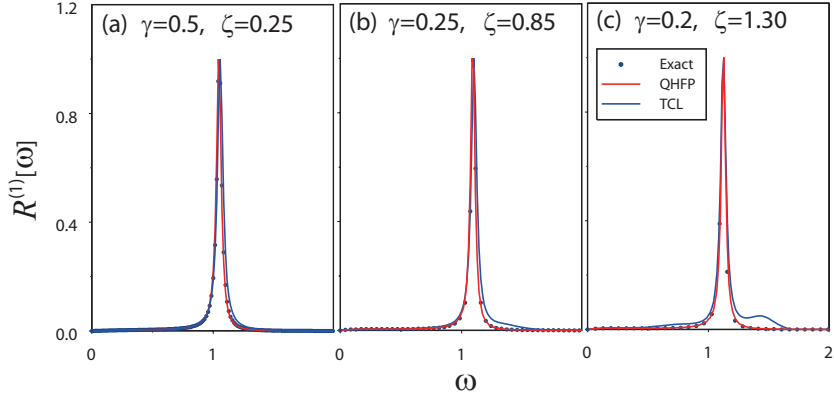}}
\end{center} 
\caption{\label{CAFigGamma}The pure non-Markovian effect of $R^{(1)}[\omega]$  investigated in the weak system-bath coupling regime. Because the effective coupling strength, $\zeta_{eff} \approx \zeta \gamma^2 \omega_0/(\gamma^2 +\omega_0^2 )$, depends on $\gamma$, we adjust $\zeta$ in each case to keep $\zeta_{eff}$ equal to its value in the case considered in Fig. \ref{R1Fig}(a). We chose (a) $\gamma=0.5$ and $\zeta=0.25$, (b) $\gamma=0.25$ and $\zeta=0.85$, and (c) $\gamma=0.2$ and $\zeta=1.3$ in order to make the widths of all the peaks similar. The other parameter values are the same as in the case of Fig. \ref{R1Fig}. The dots, red solid, and blue solid curves are the exact, QHFP and TCL Redfield without the RWA results, respectively.} 
\end{figure}
\begin{figure}
\begin{center}
\scalebox{0.6}{\includegraphics{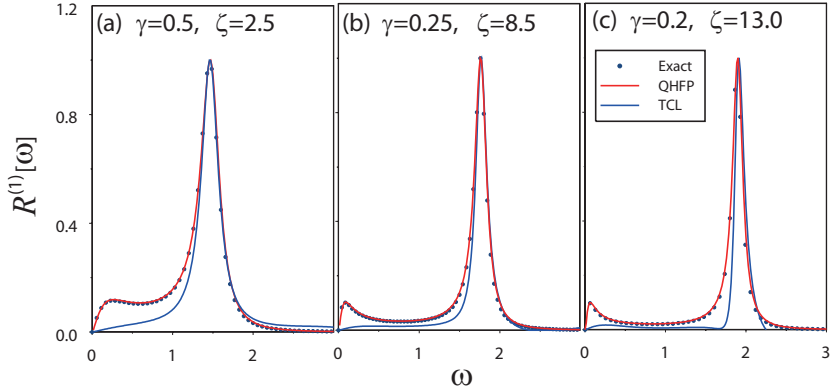}}
\end{center} 
\caption{\label{CAFigGamma2}The pure non-Markovian effect of $R^{(1)}[\omega]$ investigated in the intermediate system-bath coupling regime. We chose (a) $\gamma=0.5$ and $\zeta=2.5$, (b) $\gamma=0.25$ and $\zeta=8.5$, and (c) $\gamma=0.2$ and $\zeta=13$ in order for the effective coupling strength $\zeta_{eff} \approx \zeta \gamma^2 \omega_0/(\gamma^2 +\omega_0^2 )$ to be the same as in the case of
Fig. \ref{R1Fig}(b). The dots, red solid, and blue solid curves are the exact, QHFP and TCL Redfield without the RWA results, respectively.} 
\end{figure}

\subsubsection{Linear response function: Noise correlation and non-Markovian effects}
We next discuss the non-Markovian effects in the Brownian oscillator system. It should be noted that when the inverse noise correlation time, $\gamma$, is decreased, the effective coupling strength becomes stronger, even if we fix $\zeta$, because the bath can interact with the system multiple times when the correlation time is long. In order to study the pure non-Markovian effects, here we employ an effective coupling strength $\zeta_{eff} \approx \delta \Gamma^2(\omega_0)=\zeta \gamma^2 \omega_0/(\gamma^2 +\omega_0^2)$\cite{TanimuraJCP92} and fix it while varying $\gamma$. 

In Fig. \ref{CAFigGamma}, we plot $R^{(1)}[\omega]$ in the weak coupling regime corresponding to Fig. \ref{R1Fig}(a). Hereafter, we do not consider the TCL Redfield equation with the RWA, because the difference between the TCL results with and without the RWA is minor. While all of the peak profiles are similar if we fix $\zeta_{eff}$, the peak position shifts slightly in the high-frequency direction, because a change of $\gamma$ results in a change of $\delta \Omega^2(\omega_0)$. As the exact results and the QHFP results in Fig. \ref{CAFigGamma} indicate, there is no clear indication of non-Markovian dynamics in this weak coupling regime, once we have normalized the effective coupling strength.

While the TCL Redfield results are close to the exact results in the fast modulation (weak non-Markovian) case depicted in Fig. \ref{CAFigGamma}(a), they differ significantly in the slow modulation (strong non-Markovian) case considered in Fig. \ref{CAFigGamma}(c). This is because the perturbative description of the TCL Redfield equation breaks down as a result of the fact that multiple system-bath interactions arise due to the slow modulation, even in the weak coupling case. Thus the TCL-Redfield result without the RWA becomes negative for $\omega >4$.

In Fig. \ref{CAFigGamma2}, we plot $R^{(1)}[\omega]$ in the intermediate coupling regime corresponding to Fig. \ref{R1Fig}(b). It is seen that while the QHFP results always coincide with the exact results, the discrepancy between the TCL Redfield and exact results is large in the slow modulation (strong non-Markovian) case, due to the non-perturbative nature of the interactions. Specifically, the lack of a non-Markovian bosonic peak becomes apparent even at this intermediate coupling strength if the modulation is slow. Moreover, the TCL result without the RWA becomes negative in the region $\omega >2.2$. 
Because the non-Markovian effects in dynamics make the non-perturbative nature of the interaction conspicuous in the case of slower modulation, due to the existence of the multiple system-bath interactions for slow modulation, the TCL Redfield equation does not have the capability of treating pure non-Markovian effects even it reproduces the high-frequency part reasonably well.

\begin{figure}
\begin{center}
\scalebox{0.7}{\includegraphics{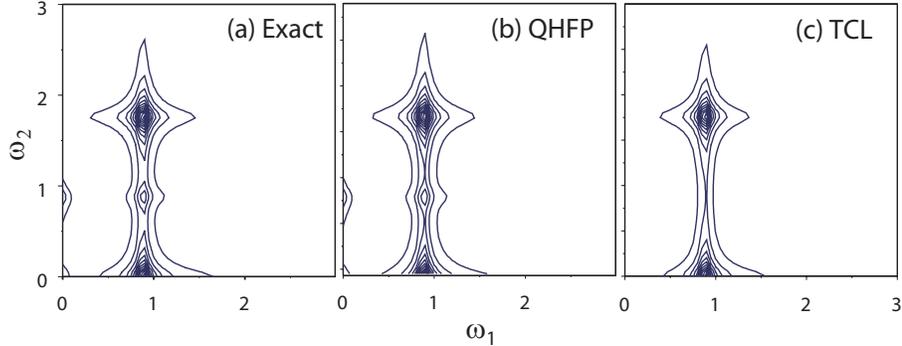}}
\end{center} 
\caption{\label{FFT2Dmiddle}The second-order response function $R^{(2)}_{TTR}[\omega_1, \omega_2]$ of the Brownian oscillator system corresponding to the intermediate coupling case considered in Fig. \ref{CAFigGamma2} (b). The results here were obtained from (a) the analytical expression Eq.\eqref{eq:2DTTR}, (b) the QHFP approach, and (c) the TCL Redfield approach without the RWA. The intensity of each graph is normalized with respect to the maximum peak intensity.} 
\end{figure}

\subsection{Nonlinear response function: Dynamical system-bath coherence}
As explained in Appendix C, the system-bath interaction induces static effects arising in imaginary time and dynamic effects arising in real time and complex time. While the static effects can be obtained from the thermal equilibrium distribution, as illustrated in Sec. V-A, we have to study the nonlinear response function to elucidate the dynamic effects. 
It should be noted that, in addition to their inability to treat strongly non-Markovian dynamics, the conventional reduced equation of motion approaches involving the TCL Redfield equation have a severe limitation in studying systems subject to time-dependent external forces because their description of the damping kernels is based on energy eigenstates.\cite{IshizakiCP08} The capability of an approach to treat external forces can also be examined by calculating nonlinear response functions, because nonlinear measurements can capture the effects of multiple interactions through time-dependent external forces. 
Here, we calculate the second-order nonlinear response function of the position given by
\begin{eqnarray}
R_{TTR}^{(2)}(t_1,t_2)=- \frac{1}{\hbar^2}\langle[[\hat q^2 (t_1+t_2), \hat q(t_1)], \hat q] \rangle.
\label{eq:R_TTR}
\end{eqnarray}
This is an observable in two-dimensional THz-Raman spectroscopy system.\cite{2drt2, Ikeda2015} Note that, because of the Gaussian integral involved in the expectation value ($\langle  \cdot \cdot \cdot \rangle = tr \{  \cdot  \cdot \cdot \exp({-\beta \hat H_{tot}}) \}$), the contribution from the lowest-order response, $\langle[[\hat q (t_1+t_2), \hat q(t_1)], \hat q] \rangle/\hbar^2$, vanishes.\cite{TanimuraJPSJ06, Ikeda2015} In the harmonic case, there is also a contribution from $R_{TRT}^{(2)}(t_1,t_2)=-\langle[[\hat q(t_1+t_2), \hat q^2(t_1)], \hat q] \rangle/\hbar^2$, which corresponds to an observable in 2D THz-Raman-THz spectroscopy system. We find that to explore the system-bath coherence, Eq.\eqref{eq:R_TTR} is suitable, as we show below.
This response function in the harmonic Brownian case can be calculated analytically as\cite{2dcl1}
\begin{eqnarray}
R_{TTR}^{(2)}(t_1,t_2)=C(t_2)C(t_1+t_2),
\label{eq:2DTTR}
\end{eqnarray}
where $C(t)$ is obtained from the Fourier transform of Eq.\eqref{eq:C_w}.  
To apply the Liouville operator formalism, we rewrite Eq.\eqref{eq:R_TTR} as
\begin{eqnarray}
R_{TTR}^{(2)}(t_1,t_2)=-\frac{1}{\hbar^2} tr \left\{ \hat q^2 \hat G(t_2)\hat q^{\times} \hat G(t_1) \hat q^{\times} \hat \rho_{tot}^{eq}\right\}.
\label{eq:R2_L}
\end{eqnarray}
Using the above expression, we calculated $R_{TTR}^{(2)}(t_1,t_2)$ for various values of $t_1$ and $t_2$ by extending the method employing Eqs. \eqref{eq:C_L} and \eqref{eq:R1_L}.\cite{TanimuraJPSJ06,TanimuraCP98} The response functions evaluated from Eqs. \eqref{eq:2DTTR} and \eqref{eq:R2_L} are then Fourier transformed to obtain two-dimensional spectra, $R_{TTR}^{(2)}[\omega_1,\omega_2]$.

In Fig. \ref{FFT2Dmiddle}, we plot 2D spectra in the frequency domain obtained from (a) the analytically exact approach, (b) the QHFP approach, and (c) the TCL Redfield without the RWA approach under the same physical conditions as in Fig. \ref{CAFigGamma2} (b). We find that while the analytically exact and QHFP results exhibit peaks at $(\omega_1, \omega_2)=(0,1)$ and $(\omega_1, \omega_2)=(1,1)$, the TCL approach cannot reproduce them. As shown in a study of multi-dimensional spectroscopy, in order to have these peaks, the dynamical system-bath coherence subject to the second interaction at time $t_1$ must be maintained throughout the time evolution described by Eq.\eqref{eq:R2_L}.\cite{ IshizakiCP08} 
In the TCL case, however, the time evolution is described in terms of the reduced operator $ tr_B \{\hat G(t) \}$, derived from the factorization assumption with $ tr \left\{ \hat q^2 tr_B \{\hat G(t_2) \} \hat q^{\times} tr_B \{ \hat G(t_1) \} \hat q^{\times} tr_B \{ \hat \rho_{tot}^{eq} \} \right\}$. 
While the exact dynamics maintain the coherence during the period of length  $t_1+t_2$ expressed by $C(t_1+t_2)$, the TCL approach cannot maintain this coherence. In contrast to the Redfield approach, because the HEOM approach can store this coherence in the hierarchal members, it is capable of treating a nonlinear response function.

Because many modern experiments utilize the nonlinear response of a system, which is measured by applying a variety of time-dependent external forces, the capability to calculate the nonlinear response function is important. The validity of the HEOM approach has been demonstrated for systems subject to time-dependent external forces.\cite{TanimuraMukamelJPSJ94, TanimuraJCP92,TanimuraMaruyama97, MaruyamaTanimura98,KatoJPCB13}
In addition to the HEOM approach, the path integral approach has also been shown to have this capability.\cite{Makri2D2010}

\begin{figure}
\begin{center}
\scalebox{0.5}{\includegraphics{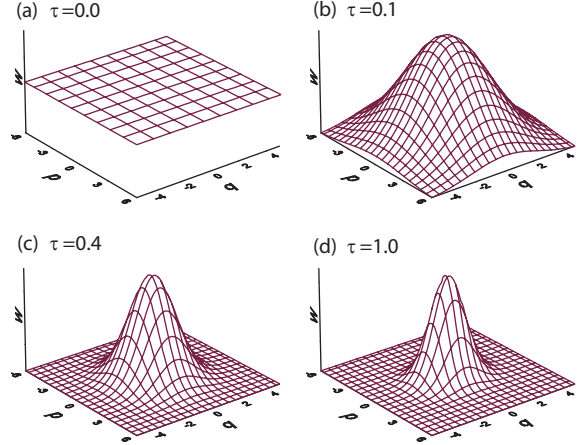}}
\end{center} 
\caption{\label{imWigner}Solution of the imaginary HEOM at four values of the imaginary time, $\tau$. Here, we plot the zeroth member, ${\bar W}^{[0,0]}(p,q; \tau)$, only. The initial state is presented in (a) $\tau=0$, while the final state is presented in (d). We confirmed that the normalized distribution of the state in (d) is identical to the distribution given by Eq.\eqref{eq:WeqBO}, within numerical error.}
\end{figure}

\begin{figure}
\begin{center}
\scalebox{0.4}{\includegraphics{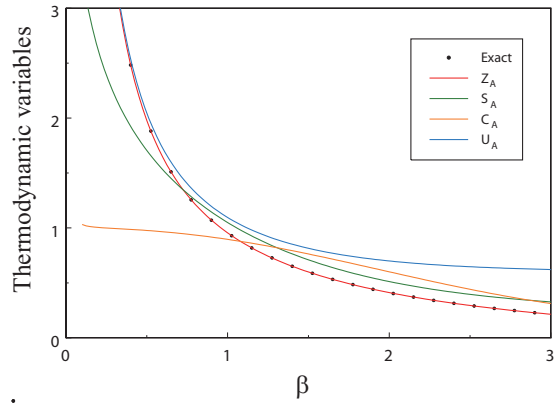}}
\end{center} 
\caption{\label{ZFUC}The partition function, $Z_A$, entropy, $S_A$, internal energy, $U_A$, and heat capacity, $C_A$, of a Brownian oscillator system calculated using the imaginary-time QHFP as functions of the inverse temperature, $\beta\hbar$. The dotted curve represent the partition function obtained from the analytical expression, Eq.\eqref{eq:ZBO_A}. Because analytically calculated exact results and the HEOM results, $Z_{\rm A}$, are nearly identical, here we plot only $S_A$, $U_A$, and $C_A$ for the HEOM case.}
\end{figure}

\subsection{Thermal equilibrium state and thermodynamic quantities}
We finally examine the imaginary QHFP equation by considering our results obtained through numerical integration of Eq.\eqref{eq:ImHEOMWig} from $\tau=0$ to $\beta \hbar$ using the harmonic potential to compare ${W}_{\rm BO}^{eq}(p, q)$ presented in V-A and the partition function $Z_{A}$. The number of Matsubara frequencies used in the imaginary-time QHFP is $K=4$. The mesh size was optimized for the Euclidean Liouvillian, and we chose $n_p=60\text{--}120$ and $n_q =120\text{--}240$. Because the distribution is spread relatively widely in the higher temperature case, we employed a coarser mesh in that case.

In Fig.\ref{imWigner}, we display solution of the imaginary-time QHFP, Eq.\eqref{eq:ImHEOMWig} with $\beta\hbar=1$ for several values of $\tau$. Because the damping kernels in the imaginary-time QHFP are defined by the Matsubara frequency at $\beta\hbar$, the solutions $\tau<\beta\hbar$ do not correspond to the equilibrium distribution at temperature $\tau$. While the initial distribution is flat, the distribution approaches a Gaussian form due to the Euclidean and the damping operators. At $\tau=\beta \hbar$, the solution coincides with the analytical solution given in Eq.\eqref{eq:WeqBO} with Eqs.\eqref{eq:QQBO} and \eqref{eq:PPBO}. 

While the equilibrium distribution can also be obtained from the real-time QHFP, as shown in Sec. V-A, the thermodynamic quantities can only be calculated from the imaginary-time QHFP. We next demonstrate this point. 
In the BO case, the partition function can also be evaluated analytically in terms of the Matsubara frequencies as\cite{GrabertZPhys84, GrabertPR88, Weiss08}
\begin{eqnarray}
Z_{\rm BO} = \frac{1}{\beta\hbar\omega_0} \prod_{k=1}^{\infty} \frac{\nu_k^2}{\omega_0^2+\nu_k^2+\delta \Gamma^2 (\nu_k)}.
\label{eq:ZBO_A}
\end{eqnarray}
We should note that, the normalization constant of the real-time QHFP is $\bar N=2\pi\sqrt{\langle p^2 \rangle \langle q^2 \rangle}$, whereas that of the imaginary-time QHFP is $Z_A$ obtained from Eq.\eqref{eq:ZHEOM}. Because $Z_{\rm BO}$ involves a temperature dependent factor other than $\bar N$, we cannot calculate the partition function using the real-time HEOM approach.

To obtain thermodynamic quantities, we first repeated the integration of the imaginary-time QHFP from $\beta\hbar=0.025$ to $3.05$ with step size $\Delta\beta\hbar =0.025$ to derive $Z_{A}$. Then, we calculated thermodynamic quantities through $Z_{A}$. In Fig. \ref{ZFUC}, we compare the partition function given by Eq. \eqref{eq:ZHEOM} (brown curve) and that obtained from Eq. \eqref{eq:ZBO_A} (dotted curve).  As expected, the imaginary-time QHFP results coincide with the exact results. For the purpose of demonstration, we also plot the entropy, $S_A$, the internal energy, $U_A$, and the heat capacity, $C_A$, calculated with the imaginary-time QHFP.
The behavior in the high temperature regime is very different from that in the spin-Boson case,\cite{Tanimura2014} because the BO model has an infinite number of excited states.

\section{Concluding Remarks}
In this paper, we presented real-time and imaginary-time QHFP equations derived using the influence functional formalism with correlated initial conditions. While we found that the QHFP equations in real time possess the same form as those obtained from a factorized initial state, we introduced a modified terminator in order to facilitate the more efficient calculations of non-Markovian dynamics.

The capability of the real-time QHFP was verified through non-perturbative and non-Markovian tests based on (i) the steady-state distribution, (ii) the symmetric auto-correlation function, (iii) the linear response function, and (iv) the nonlinear response function. This was done to test the capability of the real-time QHFP to properly model the effects of (i) static system-bath coherence, (ii) fluctuation, (iii) dissipation and non-Markovian effects, and (iv) dynamical system-bath coherence, respectively. The ability of the model to account for the dynamical system-bath coherence is particularly important if we wish to study dynamics under time dependent external forces. While many of the methodologies developed for reduced quantum dynamics have been tested only with regards to the relaxation dynamics of the population state over short periods of time, the long-time behavior of the dynamics, represented by the low frequency parts of the correlation functions, is essential to test the capability of this approach for non-Markovian dynamics. Because the bath can interact with the system many times in the case of slow modulation, the dynamics of the reduced system can only be described with a non-perturbative treatment when the system is strongly in non-Markovian. For this reason, the non-perturbative treatment and the mixed state (or unfactorized) treatment of the system-bath interactions are both important. 


In this paper, we considered only the harmonic case, the HEOM approach can be used to treat potentials of any form with time-dependent external forces. Although it had not been shown until the present paper that the QHFP equations derived from correlated initial conditions have the same form as those obtained from factorized initial conditions, the usefulness of the real-time QHFP approach has been demonstrated for various problems involving chemical reactions, \cite{TanimuraPRA91,TanimuraJCP92} photo-dissociation, \cite{TanimuraMaruyama97, MaruyamaTanimura98}
nonlinear optical response, \cite{SteffenTanimura00,TanimuraSteffen00,KatoTanimura02,KatoTanimura04, 
SakuraiJPC11} resonant tunneling,\cite{SakuraiJPSJ13, SakuraiNJP14} quantum ratchets,\cite{KatoJPCB13} and tightly bound electron-phonon system.\cite{YaoYaoJCP14} However, with the modified terminator introduced in this paper, the same calculations can be carried out more efficiently.

A confined potential system involving a Brownian oscillator system can also be treated using the HEOM approach in the energy eigenstate representation\cite{Tanimura2014} in the same manner as in the present study of the TCL Redfield equation, but quantum transport problems characterized by open or periodic boundary conditions can be studied only with the QHFP approach, \cite{TanimuraJCP92, TanimuraMaruyama97, MaruyamaTanimura98,SakuraiJPSJ13, SakuraiNJP14,KatoJPCB13} because we cannot introduce the energy eigenstates for this kind of problem. Nonlinear system-bath coupling, which plays an important role in vibrational spectroscopy, can also be taken into account in the QHFP formalism.\cite{TanimuraJPSJ06,SteffenTanimura00,TanimuraSteffen00,KatoTanimura02,KatoTanimura04,SakuraiJPC11} Extension to multi-potential surfaces is also possible.\cite{TanimuraMaruyama97, MaruyamaTanimura98} Because this formalism treats the quantum and classical systems with any form of potential from the same point of view, it allows identification of purely quantum mechanical effects through comparison of classical and quantum results in the Wigner distribution.\cite{TanimuraJCP92,SakuraiJPC11,KatoJPCB13} 

We showed that the thermal equilibrium state obtained from the imaginary-time QHFP is equivalent to the steady state solution of the real-time QHFP. Because the imaginary-time QHFP is defined in terms of integrals carried out over the definite time interval, we were able to calculate the equilibrium state more easily in this case than in the case of the real-time QHFP. Moreover, using the imaginary-time QHFP, we were able to calculate the partition function, and from this, we could directly obtain several thermodynamic quantities, namely, the free energy, entropy, internal energy, and heat capacity of the system in the dissipative environment. 
Numerical integration of the real-time and imaginary-time QHFP equations is computationally intensive. Nevertheless, we were able to study the dynamics of one-dimensional potential systems using personal computers. \cite{SakuraiJPC11,SakuraiJPSJ13,SakuraiNJP14,KatoJPCB13} Great effort has been made to reduce the computational intensiveness of algorithms used to implement the real-time HEOM approach. For example, the hierarchy has been optimized for numerical calculations,\cite{Shi09,YanPade10A,YanPade10B,Aspru11, Cao2013,Shi2013, Nike2015} and a graphic processing unit (GPU)\cite{GPU11} and parallel computers\cite{ Schuten12} have been utilized in order to facilitate the treatment of larger systems and to treat non-Drude type spectral distribution functions.\cite{TanimuraMukamelJPSJ94,TanakaJPSJ09,TanakaJCP10,TanimruaJCP12,Nori12,YanBO12,Shi14,KramerFMO,KramerJPC2013} The same techniques can be applied to the case of real-time and imaginary-time QHFP equations.

As supplementary materials, we supply the FORTRAN codes for the real-time and imaginary-time QHFP, entitled TanimuranFP15 and ImTanimuranFP15, to help further development in this field.\cite{tanimuranFP15}

\begin{acknowledgments}
Financial support from a Grant-in-Aid for Scientific Research (A26248005) from the Japan Society for the Promotion of Science is acknowledged. 
\end{acknowledgments}

\appendix

\section{Influence functional with correlated initial conditions}
The reduced density matrix elements of the system are obtained in path integral form as
\begin{align}
\rho(q,q';t) =&  \frac{ Z_B }{Z_{tot}}\int_{q_0=q(0)}^{q=q(t)} D[q(t)] \int_{q_0'=q' (0)}^{q'=q' (t)} D[q'(t)] \int_{q_0=\bar q(0)}^{q_0'=\bar q(\beta \hbar)} D[\bar q(\tau)] 
\nonumber \\
&\times {\rm e}^{\frac{i}{\hbar}S_A[q,t]} 
{\rm e}^{ \bar \Phi[q,
 q', \bar q;\; t, \beta\hbar]} 
{\rm e}^{ -\frac1{\hbar} \bar S_A[\bar q; \tau]}
{\rm e}^{-\frac{i}{\hbar}S_A[q',t]},
\label{rho_F}
\end{align}
where 
$Z_B$ is the partition function of the bath and the Euclidean action is given by
\begin{eqnarray}
 \bar S_A[\bar q; \tau] = \int_{0}^{\beta\hbar} d \tau \left\{ \frac{1}{2}m {\dot {\bar q}(\tau)}^2 + U({\bar q}(\tau))  \right\}.
\end{eqnarray}
The influence functional for the correlated initial state expressed in terms of the influence phase is given by\cite{Tanimura2014}
\begin{align}
\bar \Phi[q, q', \bar q;\; t, \beta\hbar]&= 
 {  \left( -\frac{i}{\hbar}  \right)^2\int_{{0}}^t {dt'' } \frac{i\hbar}{2} B (0)}
 {q^ \times }(t'')q^{\circ} (t'')  \nonumber \\ 
&+  \left( -\frac{i}{\hbar}  \right)^2 \int_{0}^t {dt'' } \int_{0}^{t''} {dt' } \,
q^{\times}(t'') 
 \left[-iL_1 (t'' - t')
q^{\circ}(t') 
  + L_2(t''  - t' )
q^{\times}(t')  \right]  \nonumber \\ 
&+  \frac{i}{\hbar^2} 
\int_{0}^{t} d t'' \int_{0}^{\beta\hbar} d \tau'  q^{\times}(t'') \bar  q(\tau') L (t'' +i \tau' ) \nonumber \\
&- \frac1{2\hbar^2} \int_{0}^{\beta\hbar} d \tau''B(0) \bar q^2(\tau'') 
+ \frac1{\hbar^2} 
\int_{0}^{\beta\hbar} d \tau'' \int_{0}^{\tau''} d  \tau'  \bar q(\tau'') \bar q( \tau') \bar  L \left(\tau'' -\tau'  \right),
\label{Forg}
\end{align}
where $q^\times(t) \equiv q(t) - q'(t)$ and $q^\circ(t) \equiv q(t)+q'(t)$.
Using the spectral density, $J(\omega)$, 
we rewrite these functions for $0<\tau<\beta\hbar$ as
\begin{eqnarray}
L(t+i \tau) =   \frac{2}{\beta\hbar}\int_{0}^{\infty}  d\omega J(\omega)  
\left[\frac1{\omega} + \sum_{k=1}^{\infty} \frac{2\omega}{\nu_k^2+\omega^2}  \cos (\nu_k \tau)  \right] \cos(\omega t) \\ \nonumber
 +i  \frac{2}{\beta\hbar} \int_{0}^{\infty}  d\omega J(\omega)  
 \sum_{k=1}^{\infty} \frac{2\nu_k}{\nu_k^2+\omega^2}  \sin (\nu_k \tau)  \sin(\omega t). 
\label{eq:LFourier}
\end{eqnarray}
In the case $\tau=0$, we have $L(t)\equiv iL_1(t)+ L_2(t)$ with
\begin{equation}
L_1(t) =  \int_0^{\infty} d\omega J(\omega) \sin(\omega t),
\end{equation}
\begin{align}
L_2(t) 
=\int_0^{\infty}  d\omega J(\omega) \coth \left( \frac{\beta \hbar \omega}{2} \right)\cos(\omega t),
\end{align}
and in the case $t=0$, we have $\bar L(\tau) \equiv L(i \tau)$ with
\begin{eqnarray}
\bar L(\tau) =   \frac{2}{\beta\hbar} \int_0^{\infty}  d\omega J(\omega) \left[\frac1{\omega} + \sum_{k=1}^{\infty} \frac{2\omega}{\nu_k^2+\omega^2} \cos(\nu_k \tau)
\label{eq:Lbardef} \right],
\end{eqnarray}
where the quantities $\nu_k \equiv 2\pi k/\beta \hbar$ are the Matsubara frequencies. For later convenience, we also introduce the canonical correlation 
\begin{equation}
B(t) = \frac{2}{\hbar}\int_0^{\infty}  d\omega \frac{J(\omega)}{\omega} \cos(\omega t),
\label{eq:B_t} 
\end{equation}
and express the counter-term of the potential using $B(0)$.

The function $L_2(t)$ is related to $B(t)$ through the quantum version of the 
fluctuation-dissipation theorem,
$L_2[\omega] = \hbar \omega \coth(\beta \hbar \omega/2) B[\omega]/2 $,
which insures that the system evolves toward the thermal equilibrium state, ${\rm{tr_B}} \{ \exp[-\beta \hat{H}_{tot}] \}$, for finite temperatures in the case that there is no driving force.\cite{KuboToda85} 

\begin{figure}
\begin{center}
\scalebox{0.3}{\includegraphics{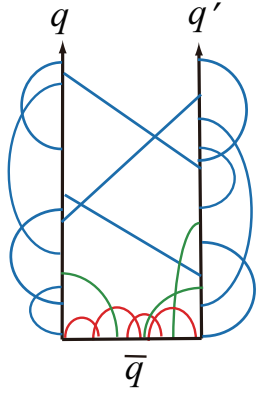}}
\end{center} 
\caption{\label{LFig}The roles of the system-bath interactions illustrated schematically. The solid black lines represent the wave function of the system A along the complex counter-path (see Fig. 1 in Ref. \onlinecite{Tanimura2014}), and the arcs and blue lines represent the system-bath interactions. Because the bath degrees of freedom have been reduced, the bath interactions connect the wave function of the system A at multiple complex times.
The blue arcs and lines correspond to the fluctuation and dissipation processes described by the terms containing $B(t''-t')$ and $L_2(t''-t')$ in Eq.\eqref{if}, while the red arcs represent the static thermal system-bath correlation described by $\bar L(\tau''-\tau')$ in Eq.\eqref{eq:PCI}. The green arcs represent the correlation in complex time described by $L(t'+\tau')$, which leads to the correlated initial conditions.}
\end{figure}

Using the relations
\begin{align}
&-\int_{{0}}^t {dt''} B (0){q^ \times }( t''){q^\circ }( t'')+
\int_{{0}}^t {d t''} \int_{{0}}^{ t''}{d t' } \;\frac{{dB (t'' - t' )}}{{d t' }}{q^ \times }( t''){q^\circ }( t' ) \nonumber \\
 &= - \int_{{0}}^t {d t''} B (t''){q^ \times }( t''){q^\circ }({0}) 
 - \int_{{0}}^t {d t''} \int_{{0}}^{ t''} {d t'} \;B (t'' - t' ){q^ \times }( t'')\frac{{d{q^\circ }( t' )}}{{d t' }}
\end{align}
and
\begin{align}
\frac1{\hbar^2}\int_{0}^{\beta\hbar} d \tau'' \int_{0}^{\tau''} d  \tau'  \bar q(\tau'') \bar q( \tau') \bar  L \left(\tau'' -\tau'  \right) - \frac1{2\hbar^2} \int_{0}^{\beta\hbar} d \tau''B(0) \bar q^2(\tau'')  \nonumber \\
=\frac1{2\hbar^2} 
\int_{0}^{\beta\hbar} d \tau'' \int_{0}^{\beta\hbar} d  \tau'  \bar q(\tau'') \bar q( \tau') \bar  L \left(\tau'' -\tau'  \right),
\end{align}
the influence functional can be rewritten as
\begin{align}
F_{CI}[q, q', \bar q;\; t, \beta\hbar]&=
{\rm e}^{ \left( -\frac{i}{\hbar}  \right)^2\int_{{0}}^t {d t''} 
\frac{{i\hbar }}{2}B (t''){q^ \times }( t''){q^\circ }(0)} 
\nonumber \\
&\times {\rm e}^{  \left( -\frac{i}{\hbar}  \right)^2  \int_{0}^t {d t'' } 
q^{\times}( t'') 
\int_{0}^{ t''} {d t' } \,\frac{i\hbar}{2} B (t'' - t')
\frac{\partial q^{\circ}( t')}{\partial t' } } \nonumber \\
&\times {\rm e}^{  \left( -\frac{i}{\hbar}  \right)^2  \int_{0}^t {d
 t'' } q^{\times}( t'') 
\int_{0}^{ t''} {d t'} L_2(t''  - t' )
q^{\times}( t')} \nonumber \\
&\times {\rm e}^{  \frac{i}{\hbar^2} 
\int_{0}^{t} d t'' \int_{0}^{\beta\hbar} d \tau'  q^{\times}(t'') \bar  q(\tau')
 L (t'' +i \tau' )},
\label{if}
\end{align}
where we have included the bath part in the initial thermal state of the system $\bar \rho_A ^{eq} [\bar q ; \beta\hbar]$ as
\begin{align}
\bar \rho^{eq} [\bar q; \beta \hbar]& = Z_B
{\rm e}^{ -\frac1{\hbar} \bar S_A[\bar q; \tau]
+ \frac1{2\hbar^2} 
\int_{0}^{\beta\hbar} d \tau'' \int_{0}^{\beta\hbar} d  \tau'  \bar q(\tau'')
\bar q( \tau') \bar  L \left(\tau'' -\tau'  \right) }.  
\label{eq:PCI}
\end{align}
The contributions arising from the factorized initial conditions or correlated initial conditions consist of two parts. One is a static contribution represented by the term containing the imaginary-time integrals of $\bar  L \left(\tau'' -\tau'  \right)$ in Eq. \eqref{eq:PCI}. Because of this term, the thermal equilibrium state of the system is not the equilibrium state of the system alone (pure state), but that of the combination of the system and bath (mixed state). The other is the correlated state contribution, represented by the term containing the complex time integrals of $L(t' +i \tau' )$ in Eq.\eqref{if}. 
The second contribution involves the effects of the dynamical correlation and is negligible when the Markovian assumption is applied, while the first contribution always plays a significant role. It is important to note that, in addition to the fluctuation and dissipation denoted by $L_2(t)$ and $B(t)$, respectively, there is a dynamical contribution from the correlated initial conditions.
The role of the system-bath interaction is illustrated in terms of each contribution in Fig. \ref{LFig}.

\begin{figure}
\begin{center}
\scalebox{0.6}{\includegraphics{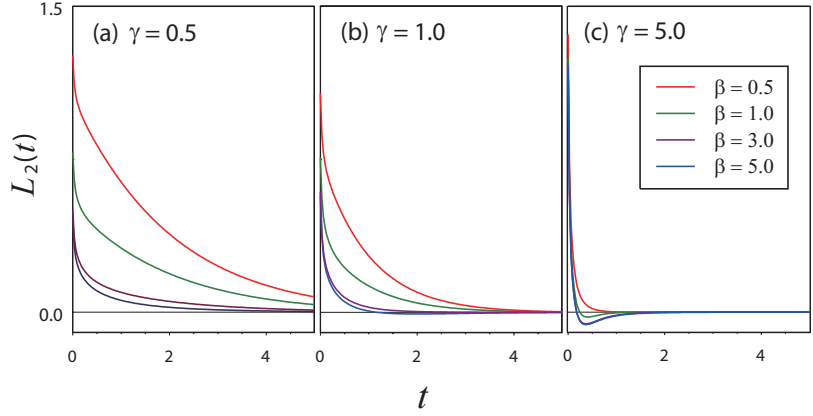}}
\end{center} 
\caption{\label{L_2Fig}The noise correlation function, $L_2(t)$, depicted as a function of the dimensionless time $t$ for several values of the inverse noise correlation time: (a) $\gamma= 0.5$, (b) $\gamma=1$,(c) $\gamma =5$.
Note that $\gamma \to \infty$ corresponds to the Markovian (Ohmic) limit,
as can be seen from Eq \eqref{JDrude}.
The inverse temperatures are, from top to bottom, $\beta \hbar =$ 0.5, 1.0, 3.0, and 5. The noise correlation becomes negative in (b) and (c) at low temperature (large $\beta\hbar$ ) due to the contribution of the Matsubara frequency terms.}
\end{figure}

\section{Drude spectral distribution and the violation of the positivity condition}
With Eq.\eqref{JDrude}, for $0<\tau<\beta\hbar$, we obtain\cite{Tanimura2014}
\begin{eqnarray}
L(t+i \tau) = &&
\left\{ c_0''  +\sum_{k=1}^{\infty}    \left[c_k''\cos (\nu_k \tau ) + i c_k'  \sin (\nu_k \tau ) \right] \right\} e^{-\gamma t} \nonumber \\
&&  +  \sum_{k=1}^{\infty} c_k' \left[\cos (\nu_k \tau ) -i \sin (\nu_k \tau ) \right]
 e^{-\nu_k t}, 
\label{eq:Ltitau2}
\end{eqnarray}
where $c_0'' = { m\zeta \gamma }/{\beta}$,
$c_k'  = -{2 m\zeta \gamma ^2 \nu_k }/{\beta (\gamma^2  - \nu_k^2 )}$, and
$ c_k''  = {2 m\zeta \gamma ^3 }/\beta (\gamma^2  - \nu_k^2)$ for $k\le0$.
At $t=0$, the above equation reduces to
\begin{eqnarray}
\bar L(\tau) =  \sum_{k=0}^{\infty}  \bar c_k \cos (\nu_k \tau),
\label{eq:LbarD}
\end{eqnarray}
where $\nu_0=0$, $\bar c_0=c_0''$, and  $\bar c_k = c_k'  + c_k''$ for $1 \le k$, while at $\tau=0$, we have
\begin{equation}
B(t) = {m \zeta \gamma}{\rm e}^{ - \gamma  t }
\label{eq:B}
\end{equation}
and
\begin{eqnarray}
 L_2(t) = c_0' e ^{ - \gamma  t }  + \sum\limits_{k = 1}^\infty  {c_k' } e ^{ - \nu _k t}.
\label{eq:L_2GMDef}
\end{eqnarray}
As shown in Fig. \ref{L_2Fig}, the noise correlation, $L_2(t)$, becomes negative at low temperature. This results from the contribution of the terms with $\nu_k = 2\pi k/\beta \hbar$ in the region of small $t$. This behavior is characteristic of quantum noise.\cite{TanimuraJPSJ06} 
We note that the characteristic time scale over which we have $L_2(t) < 0$ is determined by the temperature
and is not influenced by the spectral distribution $J(\omega)$.
Thus, the validity of the Markovian (or $\delta (t)$-correlated) noise assumption is limited in the quantum case to the high temperature regime. 
Approaches employing the Markovian master equation and the Redfield equation, which are usually applied to systems possessing discretized energy states, ignore or simplify such non-Markovian contributions of the fluctuation, and this is the reason that the positivity condition of the population states is broken. \cite{Davies76,Gorini78,Spohn80,Dumcke79,Pechukas94,Romero04, Frigerio81, Frigerio85} 

As a method to resolve this problem, the rotating wave approximation
(RWA) (also known as the ''secular approximation'') is often employed, but a
system treated under this approximation will not satisfy the fluctuation-dissipation theorem, and thus the use of such an approximation may introduce significant error in the thermal equilibrium state and in the time evolution of the system toward equilibrium. Because the origin of the positivity problem lies in the unphysical Markovian assumption for the fluctuation term, the situation is better in the non-Markovian case, even within the framework of the Redfield equation without the RWA, as discussed in Sec. V. In the classical limit, with $\hbar$ tending to zero, $L_2(t)$ is always positive.

While conventional approaches employing reduced equations of motion eliminate the bath degrees of freedom completely, the HEOM approach retains information with regard to the system-bath coherence in the hierarchy elements. Because of this feature, the HEOM approach can treat the reduced dynamics in a non-perturbative, non-Markovian manner. To obtain a more compact form for the HEOM, we use the following approximate form for $ L_2(t)$, given in Eq. \eqref{eq:L_2GMDef}:
$L_2(t) \simeq c_0' e ^{ - \gamma t}  + \sum_{k = 1}^{K}  {c_k' } e ^{ - \nu _k t} 
+  \delta(t) \sum_{k = K+1}^{\infty} C_k {c_k'}/{\nu_k},$ 
with $c_0'  
= \hbar m\zeta\gamma^2 \cot (\beta\hbar\gamma/2)/2$.
Here, we choose $K$ so as to satisfy $\nu_k=2\pi K/(\beta\hbar) \gg \omega_c$, where $\omega_c$ represents the characteristic frequency of the system. Under this condition, we can apply the approximation $\nu_k{\rm e}^{-\nu_k |t|}\simeq C_k \delta( t ) \quad ({\rm for} \ \ k \geq  K+1)$ with negligible error at the desired temperature, $1/\beta$, where $ C_k = {\nu_k^2}/({\nu_k^2+\omega_c^2})$ is the correction factor that compensates for the overestimation of $L_2(t)$ in the approximation 
at very low temperature for small cut-off $K$.
The accuracy of this approximation is verified on basis of the asymptotic behavior of $L_2(t)$ as a function of $K$. 
Then, the HEOM can be obtained by considering the time derivative of Eq. (\ref{rho_red}).\cite{TanimuraJPSJ06}
When the temperature becomes high (i.e. for $\beta\hbar\gamma \ll 1$), the noise correlation function reduces to $L_2(t) \simeq m\zeta \gamma e^{-\gamma|t|}/\beta$, and hence the noise modulates the system as a Gaussian-Markovian stochastic  process.\cite{TanimuraPRA91,TanimuraJCP92}

\section{Derivation of the HEOM in configuration space}
In the present appendix, we construct the equation of motion for $\rho _{{j_1}{j_2} \cdots {j_K}}^{(n)} (q,q';\,t)$.  
In order to obtain differential equations in time, we consider the reduced density matrix elements at $t + \delta t$,
\begin{align}
\rho _{{j_1}{j_2} \cdots {j_K}}^{(n)}(q,q';\,t + \delta t)& = \frac{1}{{{A^2}}}\int {dy} \int {dy'} \int_{\bar q({0}) = {q_0}}^{q(t) = q - y} D [q(t)]\int_{\bar q'({0}) = {q_0}^\prime }^{q'(t) = q' - y'} D [q'(t)]\nonumber \\
& \times \int_{q_0=\bar q(0)}^{q_0'=\bar q(\beta \hbar)} D[\bar q(\tau)] \;
\bar \rho^{eq} [\bar q; \beta\hbar] \nonumber \\
& \times {\left\{ \operatorname{e} ^{ -\gamma (t + \delta t ) } \left[
\int_{0}^{t + \delta t } dt' 
  \operatorname{e} ^{ \gamma t' }\gamma \Theta_0 (t' ) 
 +  G_0(0) -\frac1{\hbar} \bar \Theta (\beta \hbar) \right]
 \right\}^n} \nonumber \\
& \times \prod\limits_{k = 1}^K \left\{  
  \operatorname{e} ^{ -\nu_k  (t + \delta t )} \left[
\int_{0}^{t+\delta t} dt' 
  \operatorname{e} ^{ \nu_k t'} \nu_k  \Theta_k (t' ) 
-\frac1{\hbar}  \bar \Psi_k  (\beta \hbar) \right]
\right\}^{j_k}
\nonumber \\
 &\times {{\mathop{\rm e}\nolimits} ^{ \frac{i}{\hbar }S[q,t + \delta t]}}
F_{CI}[q, q', \bar q;\; t+ \delta t, \beta\hbar]
{{\mathop{\rm e}\nolimits} ^{ - \frac{i}{\hbar }S[q',t + \delta t]}},
\label{eq:hierder1}
\end{align}
where $A$ is the normalization constant for the integrals over $y$ and $y'$, and we set $q = q(t) + y$ and $q' = q'(t) + y'$ with $q(t + \delta t) = q$ and $q'(t + \delta t) = q'$. We then expand $\rho _{{j_1}{j_2} \cdots {j_K}}^{(n)}(q,q';\,t + \delta t)$ in terms of $\delta t$ up to first order. Because $y$  and $y'$  also depend on $\delta t$, we have to expand the above equations in terms of $y$ and $y'$. In the following, we expand the components separately.

The action part can be expressed as
\begin{align}
{{\mathop{\rm e}\nolimits} ^{ \frac{i}{\hbar }S[q,t + \delta t]}} &= {{\mathop{\rm e}\nolimits} ^{  \frac{i}{\hbar }\left[ {  \frac{m}{2}{{\left( {\frac{y}{{\delta t}}} \right)}^2} - U\left( {q - y} \right)} \right]\delta t}}{{\mathop{\rm e}\nolimits} ^{  \frac{i}{\hbar }S[q - y,t]}}
\nonumber \\
& = {{\mathop{\rm e}\nolimits} ^{\frac{{im{y^2}}}{{2\hbar \delta t}}}}\left( {1 - \frac{{i\delta t}}{\hbar }U\left( q \right)} \right){{\mathop{\rm e}\nolimits} ^{  \frac{i}{\hbar }S[q - y,t]}}.
\label{eq:exp2}
\end{align}
The influence functional is evaluated as
\begin{align}
F_{CI}[q, q', \bar q;\; t+ \delta t, \beta\hbar]
 &= \left[ 
1 
- \delta t \; \Phi (t) \left\{ \operatorname{e} ^{ -\gamma (t + \delta t ) } \left[
\int_{0}^{t + \delta t } dt' 
  \operatorname{e} ^{ \gamma t' }\gamma \Theta_0 (t' ) 
 +  G_0(0) -\frac1{\hbar} \bar \Theta (\beta \hbar) \right]
 \right\} \right. \nonumber \\
&- \left. \delta t \; \Phi (t) \left\{  
  \sum_{k=1}^K \operatorname{e} ^{ -\nu_k  (t + \delta t )} \left[
\int_{0}^{t+\delta t} dt' 
  \operatorname{e} ^{ \nu_k t'} \nu_k  \Theta_k (t' ) 
-\frac1{\hbar}  \bar \Psi_k  (\beta \hbar) \right]
\right\} - \delta t \Xi' (t) \right]
\nonumber \\
&\times F_{CI}[q-y, q'- y', \bar q;\; t, \beta\hbar] .
\label{eq:hierder2}
\end{align}
In the following, we apply the Gaussian integrals
\begin{align}
\frac{1}{A}\int {dy} y{{\mathop{\rm e}\nolimits} ^{\frac{{im{y^2}}}{{2\hbar \delta t}}}} = 0
\end{align}
and
\begin{align}
\frac{1}{A}\int {dy} {y^2}{{\mathop{\rm e}\nolimits} ^{\frac{{im}}{{2\hbar \delta t}}{y^2}}} = \frac{{i\hbar }}{m}\delta t ,
\end{align}
where the normalization constant is chosen to be $A = \int {dy} \exp (im{y^2}/2\hbar \delta t)$. Gaussian integrals higher than fourth order can be ignored, because they produce contributions smaller than $o(\delta t)$.
 
With Eqs.\eqref{eq:exp2} and \eqref{eq:hierder2}, the expansion of the last term in Eq. \eqref{eq:hierder1} is completed
by the following:
\begin{align}
{{\mathop{\rm e}\nolimits} ^{ \frac{i}{\hbar }S(q - y,t)}}
F_{CI}[q-y, q'- y', \bar q;\; t, \beta\hbar]
{{\mathop{\rm e}\nolimits} ^{ - \frac{i}{\hbar }S(q' - y',t)}}
  &= \left( {1 - y\frac{\partial }{{\partial q}} - y'\frac{\partial }{{\partial q'}} + \frac{{{y^2}}}{2}\frac{{{\partial ^2}}}{{\partial {q^2}}} + \frac{{{{y'}^2}}}{2}\frac{{{\partial ^2}}}{{\partial {{q'}^2}}}} \right) \nonumber \\
&\times {{\mathop{\rm e}\nolimits} ^{  \frac{i}{\hbar }S(q,t)}} F_{CI}[q, q', \bar q;\; t, \beta\hbar]{{\mathop{\rm e}\nolimits} ^{ - \frac{i}{\hbar }S(q',t)}}.
\label{eq:expand1}
\end{align}
Then, collecting the pieces from Eqs. \eqref{eq:exp2}, \eqref{eq:hierder2} and \eqref{eq:expand1}, and keeping terms up to $o(\delta t)$, we have the following for the kinetic term of the Hamiltonian: 
\begin{align}
\int {\frac{{dy}}{A}} {{\mathop{\rm e}\nolimits} ^{\frac{{im{y^2}}}{{2\hbar \delta t}}}}\left( {1 - y\frac{\partial }{{\partial q}} + \frac{{{y^2}}}{2}\frac{{{\partial ^2}}}{{\partial {q^2}}}} \right) = 1 - \delta t\frac{i}{\hbar }\left( { - \frac{{{\hbar ^2}}}{{2m}}\frac{{{\partial ^2}}}{{\partial {q^2}}}} \right).
\label{eq:gauss1}
\end{align}

We next consider the expansion of the factor ${\{  \cdots \} ^n}$ in Eq.(\ref{eq:hierder1}), first in terms of $y$ and $y'$, and then in terms of $\delta t$. For the expansion in $y$  and $y'$ up to second-order, we have
\begin{align}
\frac{{n m \zeta \gamma }}{2}\left( {y + y' } \right) {\left\{ {\int_{{0}}^t {dt' \gamma {{\mathop{\rm e}\nolimits} ^{ - \gamma (t - t' )}}} {\Theta _0}(t') + G_0(0) -\frac1{\hbar} \bar \Theta (\beta \hbar) } \right\}^{n - 1}}. 
\end{align}
This term reduces to $\rho _{{j_1}, \ldots ,{j_K}}^{\left( {n - 1} \right)}\left( {q,q';t} \right)$, and therefore the contribution of the above to the relevant order in $\delta t$ can be expressed as
\begin{align}
\frac{{n m \zeta \gamma }}{2}\int {\frac{{dy}}{A}} \int {\frac{{dy'}}{A}} {{\mathop{\rm e}\nolimits} ^{\frac{{im{y^2}}}{{2\hbar \delta t}}}}{{\mathop{\rm e}\nolimits} ^{ - \frac{{im{{y'}^2}}}{{2\hbar \delta t}}}}\left( {1 - y\frac{\partial }{{\partial q}} - y'\frac{\partial }{{\partial q'}} + \frac{{{y^2}}}{2}\frac{{{\partial ^2}}}{{\partial {q^2}}} + \frac{{{{y'}^2}}}{2}\frac{{{\partial ^2}}}{{\partial {{q'}^2}}}} \right) \nonumber \\
\quad \quad  \times \left( y +y' \right) \rho _{{j_1}, \ldots ,{j_K}}^{\left( {n - 1} \right)}\left( {q,q';t} \right)
\end{align}
Then, integrating over $y$ and $y'$, we have
\begin{align}
- \delta t\frac{{in\hbar \zeta \gamma }}{{2}}\left[ {\frac{\partial }{{\partial q}} -\frac{\partial }{{\partial q'}} } \right]\rho _{{j_1}, \ldots ,{j_K}}^{\left( {n - 1} \right)}\left( {q,q';t} \right).
\end{align}
For the expansion of ${\{  \cdots \} ^n}$ in terms of $\delta t$, we have
\begin{align}
n{\Theta _0}(t){\left\{ {\int_{{0}}^t {dt' \gamma {{\mathop{\rm e}\nolimits} ^{ - \gamma (t - t' )}}} {\Theta _0}(t') + G_0(0) -\frac1{\hbar} \bar \Theta (\beta \hbar)} \right\}^{n - 1}}\delta t \nonumber \\
 - n\gamma {\left\{ {\int_{{0}}^t {dt' \gamma {{\mathop{\rm e}\nolimits} ^{ - \gamma (t - t' )}}} {\Theta _0}(t') + G_0(0) -\frac1{\hbar} \bar \Theta (\beta \hbar)} \right\}^n}\delta t .
\label{ntheta0}
\end{align}

We can expand the factors ${\{  \cdots \} ^{j_k}}$ in Eq.\eqref{eq:hierder1}
 similarly to the  ${\{  \cdots \} ^{n}}$ factor. We obtain
\begin{align}
 - {j_k}{\nu _k}{\left\{ { - \int_{{0}}^{t} {dt' {\nu _k}{{\mathop{\rm e}\nolimits} ^{ - {\nu _k}(t - t' )}}} {\Theta _k}(t')- \frac1{\hbar}  \bar \Psi_k  (\beta \hbar)} \right\}^{{j_k}}}\delta t \nonumber \\
 - {j_k}{\nu _k}{\Theta _k}(t){\left\{ { - \int_{{0}}^{t} {dt' {\nu _k}{{\mathop{\rm e}\nolimits} ^{ - {\nu _k}(t - t' )}}} {\Theta _k}(t') \frac1{\hbar} - \bar \Psi_k  (\beta \hbar)} \right\}^{{j_k} - 1}}\delta t .
\label{jktheta}
\end{align}
Using the definition of the hierarchy elements Eqs.\eqref{eq:rhon_jk} and \eqref{auxFV}, we obtain 
\begin{align}
\left\{ { - n\gamma \rho _{{j_1}, \ldots ,{j_K}}^{\left( n \right)}\left( {q,q';t} \right) + n{\Theta _0}(t)\rho _{{j_1}, \ldots ,{j_K}}^{\left( {n - 1} \right)}\left( {q,q';t} \right)} \right\}\delta t 
\label{n-1}
\end{align}
and
\begin{align}
\left\{{-{j_k}{\nu_k}\rho _{{j_1}, \ldots ,{j_K}}^{\left( n \right)}\left( {q,q';t} \right) - {j_k \nu_k}{\Theta _k}(t)\rho _{{j_1}, \ldots ,{j_k} - 1, \ldots ,{j_K}}^{\left( n \right)}\left( {q,q';t} \right)} \right\} \delta t ,
\label{jk-1}
\end{align}
From Eqs.\eqref{ntheta0}and \eqref{jktheta}, respectively.
Finally, substituting the results from each of the above expansions, contained
in Eqs.\eqref{eq:gauss1}, \eqref{n-1} and \eqref{jk-1}, into Eq.\eqref{eq:hierder1}, we construct the complete form for this expression to $o(\delta t)$, and from this, we obtain Eq.\eqref{eq:HEOMq}.

\section{Time-Convolutionless (TCL) Redfield Equation}
The TCL Redfield equation is the reduced equation of motion in the case of non-Markovian noise whose damping kernels are expressed in a time-convolutionless form.\cite{TCLshibata77,TCLshibata79,TCLshibata10}
The TCL Redfield equation is exact if the system Hamiltonian, $\hat H_A$, is time-independent and if $\hat H_A$ commutes with the bath interaction. However, for the BO model considered in this paper, defined by Eq.\eqref{eq:BrownianH}, the system Hamiltonian does not commute with the bath interaction. 

In order to apply the Redfield theory, we need to use the eigenstate representation of the system. For this reason, we include a counter-term in the system Hamiltonian, and consider the modified Hamiltonian, $\hat H_A' = (\hat H_A + m \zeta \gamma \hat q^2 /2)$. For the $j$th eigenenergy, $E_j'= \hbar (j +1/2) \omega_0'$, where $\omega_0' = \sqrt{\omega_0^2+\zeta \gamma}$, the eigenfunction for $\hat H_A'$ is expressed in terms of Hermite polynomials, $H_j(\cdot)$, as
\begin{eqnarray}
\psi_j (q) = \left(\frac{\alpha^2}{\pi}\right)^{\frac{1}{4}} \frac{1}{\sqrt{2^j j!}} \exp\left(- \frac{\alpha^2 q^2}{2} \right)  H_j (\alpha q),
\label{eq:neigen}
\end{eqnarray}
where $\alpha = \sqrt{m \omega_0'/\hbar}$. We denote the ket vector for $ \psi_j (q) $ by $| j \rangle $. The TCL Redfield equation for the reduced density matrix elements, ${\rho}_{jk}(t) \equiv \langle j | \hat \rho_A (t) | k \rangle$, is then given by
\begin{eqnarray}
\frac{\partial}{\partial t}{\rho}_{jk}(t) = - i \omega_{jk}'  {\rho}_{jk}(t)+ \sum_{l,m} R_{jk,lm} (t) \rho_{lm}(t), 
\label{eq:RedfieldEq}
\end{eqnarray}
where $\omega_{jk}' \equiv (j-k) \omega_0'$ and $ R_{jk,lm} (t)$ is the Redfield  tensor defined by
\begin{eqnarray}
R_{jk,lm} (t) \equiv \Gamma_{mk,jl}(t) +  \Gamma_{lj,km}^{\dag}(t) -\delta_{jm}\sum_{n}  \Gamma_{jn,nl}(t) -\delta_{jl}\sum_{n}  \Gamma_{kn,nm}^{\dag}(t),
\label{eq:Redfieldtensor}
\end{eqnarray}
with
\begin{eqnarray}
\Gamma_{jk,lm}(t) = \bar \Gamma_{jklm} \left(\frac{ \zeta \gamma^2e^{-i\frac{\beta\hbar\gamma}{2}}}{2\sin\left( \frac{ \beta \hbar\gamma }{2} \right) }\frac{1-e^{-(\gamma+i \omega_{lm}' )t}}{\gamma+i \omega_{lm}'} - \frac{2}{\beta \hbar} 
\sum_{k'=1}^{\infty} \frac{\zeta \gamma^2 \nu_{k'} }{\gamma^2-\nu_{k'}^2}
\frac{1-e^{-(\nu_{k'} + i \omega_{lm}')t}}{\nu_{k'}+i \omega_{lm}'} \right).
\label{eq:Redfieldel}
\end{eqnarray}
The interaction tensor is defined by
\begin{eqnarray}
\bar \Gamma_{jklm} \equiv\langle j | \hat q | k \rangle \langle l | \hat q | m \rangle .
\label{eq:NonRWA}
\end{eqnarray}
The rotating wave approximation (RWA) is expressed as $\hat q \hat x_j 
=\sqrt{2 \hbar/m  \omega_0'} 
(\hat a^+ + \hat a^-)(\hat b_j^- + \hat b_j^+)\approx \hat a^+ \hat b_j^- + \hat a^- \hat b_j^+$, where $\hat a^{\pm}$ and $\hat b_j^{\pm}$ are the creation and annihilation operators of the BO oscillator and the $j$th bath oscillator, respectively. For $\langle j | \hat a^+ | k \rangle=0$ and $\langle k | \hat a^- | j \rangle=0$, with $j \ne k=j+1$, the interaction tensor in RWA form is given by
\begin{eqnarray}
\bar \Gamma_{jklm}^{RWA} =  \frac{2 \hbar}{m  \omega_0'} \left(
\langle j | \hat a^+ | k \rangle \langle l | \hat a^- | m \rangle + \langle j | \hat a^- | k \rangle \langle l | \hat a^+ | m \rangle \right).
\label{eq:RWA}
\end{eqnarray}

In the Wigner representation, the eigenstate elements of the density matrix are expressed as
\begin{eqnarray}
W_{jk}(p,q) =  \frac{1}
{{2\pi \hbar }}\int_{ - \infty }^\infty  {dx} \cos\left( \frac{px}{\hbar}\right)  \psi_j \left( q - \frac{x}{2} \right) 
\psi_k \left(q + \frac{x}{2} \right).
\label{eq:Wigjk}
\end{eqnarray}
The total distribution is then given by
\begin{eqnarray}
W(p,q; t) =  \sum_{j, k =1}^{M} \rho_{jk}(t) W_{jk} (p,q),
\label{eq:WigTCL}
\end{eqnarray}
where $M$ is the number of energy eigenstates employed to solve the TCL Redfield equation. The factorized initial state is expressed as
\begin{eqnarray}
W(p,q; 0) =  \sum_{j=1}^{M} \frac{1}{Z_A'} \exp(-\beta E_j' ) W_{jj} (p,q),
\label{eq:WigTCLeq}
\end{eqnarray}
where $ Z_A'=\sum_j \exp(-\beta E_j )$. By comparing the steady state solution of the TCL Redfield equation in the Wigner representation with the analytical solution of the BO model given by $W_{\rm BO}^{eq}(p, q)$ with Eqs.\eqref{eq:QQBO} and \eqref{eq:PPBO}, we can check the accuracy of the steady-state distribution in the TCL formalism.

\end{document}